\documentclass[graybox,envcountchap]{svmult}

\usepackage{mathptmx}        
\usepackage{amsmath}
\usepackage{amssymb}
\usepackage{color}
\usepackage{helvet}          
\usepackage{courier}         
\usepackage{dirtree}

\usepackage{makeidx}        
\usepackage{graphicx}        
\usepackage{subfig}

\usepackage{multicol}        
\usepackage[bottom]{footmisc}

\usepackage{hyperref}        
\hypersetup{colorlinks=true,urlcolor=blue}

\usepackage[misc]{ifsym}

\makeindex             

\begin{document}

\title*{Regular black holes in Palatini  gravity}

\author{Gonzalo J. Olmo and Diego Rubiera-Garcia}

\institute{Gonzalo J. Olmo \at Departamento de F\'{i}sica Te\'{o}rica and IFIC, Centro Mixto Universidad de Valencia - CSIC.
Universidad de Valencia, Burjassot-46100, Valencia, Spain, \email{gonzalo.olmo@uv.es}
\and Diego Rubiera-Garcia (\Letter)  \at Departamento de F\'isica Te\'orica and IPARCOS,
	Universidad Complutense de Madrid, E-28040 Madrid, Spain, \email{drubiera@ucm.es}}

\maketitle

\abstract{ Palatini (or metric-affine) theories of gravity are characterized by having {\it a priori} independent metric and affine structures. The theories built in this framework have their field equations obtained as independent variations of the action with respect to the metric, the affine connection, and the other fields. In this Invited chapter for the edited book ``Regular Black Holes: Towards a New Paradigm of the Gravitational Collapse" (Ed. C. Bambi, Springer Singapore, expected in 2023), we consider the issue of singularity-removal in several members of a family of theoretically consistent and observationally viable subclass of them, built as contractions of the Ricci tensor with the metric (Ricci-based gravities or RBGs, for short). Several types of (spherically symmetric) solutions are considered from combinations of the gravity and matter sectors satisfying basic energy conditions, discussing the modifications to the horizons and to the innermost structure of the solutions, the main player in town being the presence in some cases of a bouncing behavior in the radial function. We use a full weaponry of criteria to test whether singularity-removal has been achieved: completeness of (null and time-like) geodesic paths, impact of unbound curvature divergences, analysis of causal contact upon congruences of geodesics, paths of observers with (bound) acceleration, and propagation of (scalar) waves. We further elaborate on the (three) main avenues by which such a regularity of the corresponding space-times is achieved, and comment on the lack of correlation between the (in)completeness of geodesics and the blow up of curvature scalars, something too frequently and too carelessly assumed in the literature on the subject. We conjecture that a larger flexibility in the underlying geometrical ingredients to formulate our gravitational theories may hold (part of) the clue to resolve black hole singularities.}

\section{Introduction to metric-affine gravity and resolution of space-time singularities}

In the construction of the theories for the gravitational field considered in this chapter, the classical paradigm views gravity as the manifestation of a dynamical space-time, i.e., a differentiable manifold $\mathcal{M}$ endowed with geometrical entities acting not only as the canvas in which events happen, but which also react in a non-trivial way to the motion of energy and observers living on it. The first element in this view - the geometrical entities - is tightly attached to the roles attributed to the main characters of the geometry, namely, the metric $g$ and the affine connection $\Gamma$. While the metric is the responsible of the causal structure of space-time and, as such, it is associated to local measurements of distances, angles, areas, volumes, etc, the affine connection is associated to free-fall, defining the notion of parallelism and entering into the covariant derivatives. As such, these two objects are conceptually and operationally independent from each other. The second aspect - the dynamical behavior - is given by the particular way the geometrical elements react (and the other way round) to the motion of the matter fields, and is given by the field equations derived from the action defining the theory. The original formulation by Einstein of his General Theory of Relativity (GR) amounts to associate the affine connection to the metric via the Christoffel symbols of the latter, thus rendering a privileged role to a particular connection - the Levi-Civita one - in order for its covariant derivative to preserve the metric, and selects the lowest-order scalar object in the action  - the curvature scalar - able to endow the theory with dynamics. This way, using the duo $\{\mathcal{M},g\}$ one finds the canonical ({\it curvature-based}) formulation of GR, whose Einstein-Hilbert action yields the dynamical (Einstein) equations of the theory.

Since these events took place, theoreticians and geometers alike have not been idle. Among their many findings, for the sake of this chapter we underline the (nowadays) well understood fact that the affine connection can be split, in general, into three pieces, namely, curvature, torsion, and non-metricity, each with its own physical interpretation: the rotation of a vector transported along a closed curve, the non-closure of parallelograms when two vectors are parallely transported along each other, and the variation of a vector's length when parallel-transported. It turns out that, when the action of the gravitational theory is built upon the lowest scalar of each of these pieces (thus switching off the other two), the resulting dynamical equations are precisely the Einstein ones \cite{Jimenez:2019woj}. Diffemorphism and Lorentz invariances are preserved, and the background solutions of each formulation of GR turn out to be the same in such a way that it is only at the level of their respective boundary terms (relevant for some applications) that they can be distinguished from each other. Such three versions are called (canonical) GR, the teleparallel equivalent of GR \cite{Aldrovandi:2013wha,Maluf:2013gaa}, and the (symmetric) teleparallel GR \cite{Nester:1998mp,Obukhov:2002tm}, respectively. Fully enlarging GR to include non-vanishing contributions of every such piece of the affine connection yields a theory called metric-affine gravity, given by a triplet $\{\mathcal{M},g,\Gamma\}$ with a priori unspecified relations between $g$ and $\Gamma$. Such a theory contains not only the three equivalent formulations of GR on its corresponding limits, but also other gravitational theories explored in the literature such as Einstein-Cartan, Weyl gravity, etc. Such a picture can be very well dubbed as {\it affinesia}\footnote{Credit to Jose Beltr\'an Jim\'enez who, as far as we know, is the person to be blamed for coining this term.}.

In view of the discussion above, in order to look for a way out of the unavoidable existence of space-time singularities within GR as guaranteed by the singularity theorems \cite{Senovilla:2014gza}, one is faced at a crossroads with (at least) two important decisions to make: i) what are the underlying geometrical elements of the theory as well as the a priori relation (if any) among them, and ii) what is the action of the theory yielding (via its field equations) dynamics to them and allowing observers to ``gravitate". In metric-affine (Palatini) theories of gravity the path taken is defined by the following choices: i) restore metric and affine connection to their original roles as independent entities, taking the latter to have both non-vanishing curvature, torsion, and non-metricity pieces, and ii) enlarge the Einstein-Hilbert action towards a new scalar action in such a way that the target theory has a number of minimum properties: diffeomorphism plus Lorentz invariances, compatibility with weak-field experiments (which means it reduces to GR in the suitable limit), and absence of extra propagating degrees of freedom (removing new scalar-like fields or additional tensorial polarizations). These constraints are conservative enough so as not to enter into conflict with those experiments that GR passes \cite{Will:2014kxa}, but at the same time furnishing the resulting theory with a larger flexibility of the geometrical architecture which can also be put to observational test \cite{Bahamonde:2021akc}. This chapter amounts to a substantive discussion of this framework and the insights it has brought to our knowledge regarding the nature of space-time singularities and how to resolve them in physically reasonable enough settings. For simplicity, and since we will we dealing with bosonic fields, which are oblivious to the torsional part of the affine connection \cite{Afonso:2017bxr}, we shall neglect it from our considerations, though some comments and references will be provided when necessary.

There are many theories that can accommodate these two requirements. In order to start our discussion, we shall start with the essentials, and progress from there to the most general family of such theories studied in detail so far, dubbed as Ricci-based gravities. Let us begin.

\section{Gravitational models}

\subsection{$f(\mathcal{R})$ gravity}

The simplest extension of GR that we will consider is given by $f(\mathcal{R})$ gravity \cite{Olmo:2011uz}, where $f$ is an arbitrary function of the curvature scalar $\mathcal{R}=g^{\mu\nu}R_{\mu\nu}(\Gamma)$, where the Ricci tensor  $R_{\mu\nu}\equiv {R^\alpha}_{\mu\alpha\nu}$ is built from the Riemann tensor
\begin{equation}
{R^\rho}_{\sigma\mu\nu}=\partial_\mu \Gamma^{\rho}_{\nu\sigma} -\partial_{\nu}\Gamma^{\rho}_{\mu\sigma}+\Gamma^{\rho}_{\mu\lambda}\Gamma^{\lambda}_{\nu\sigma} - \Gamma^{\rho}_{\nu\lambda}\Gamma^{\lambda}_{\mu\sigma}
\end{equation}
made up of an independent affine (torsionless) connection, $\Gamma \equiv \Gamma_{\mu\nu}^\lambda$, which is a priori unrelated to any metric. It is worth pointing out that this Palatini curvature scalar is different from the metric curvature scalar, $R=g^{\mu\nu}R_{\mu\nu}(g)$, where in this case the Ricci tensor $R_{\mu\nu}(g)$ is the one derived from the Christoffel symbols of the space-time metric $g_{\mu\nu}$ entering in the definition of the action. This fact makes the dynamics of Palatini $f(\mathcal{R})$ gravity to dramatically depart from its metric counterpart \cite{DeFelice:2010aj}.

\begin{svgraybox}
{\bf Projective invariance:} The Einstein-Hilbert action is invariant under a class of (projective) transformations of the form $\tilde{\Gamma}_{\mu\nu}^{\lambda}=\Gamma_{\mu\nu}^{\lambda} +\xi_{\mu}\delta^{\lambda}_{\nu}$, where $\xi_\mu$ is a 1-form vector field. However, under this transformation the Ricci tensor is not invariant, transforming as $\tilde{R}_{\mu\nu}=R_{\mu\nu}+F_{\mu\nu}$, where the tensor $F_{\mu\nu}=\partial_\mu \xi_{\nu} - \partial_{\nu}\xi_{\mu}$. Since according to this, only the symmetric part of the Ricci tensor in $R_{\mu\nu}=R_{(\mu\nu)}+R_{[\mu\nu]}$ (parenthesis and brackets denote symmetrization and anti-symmetrization, respectively) is invariant under projective transformations, gravity theories based on the full Ricci tensor are prone to developing ghost-like instabilities associated to the anti-symmetric piece \cite{BeltranJimenez:2019acz}. In order to avoid this problem, in the construction of the most general class of theories consider in this Chapter, namely, Ricci-Based Gravity theories (RBGs), we will just consider the symmetric part of the Ricci tensor $R_{(\mu\nu)}$, but remove the parenthesis for notational simplicity. This subtlety is irrelevant in the $f(\mathcal{R})$ case, because the contraction with the metric filters out the antisymmetric part.
\end{svgraybox}

Let us define its action as
\begin{equation} \label{eq:actfR}
S=\frac{1}{2\kappa^2} \int d^4x \sqrt{-g} f(\mathcal{R}) + S_m(g_{\mu\nu},\psi_m)
\end{equation}
where $\kappa^2=8\pi$ in $G=c=1$ units, $g$ is the determinant of the space-time metric $g_{\mu\nu}$, and $S_m=\int d^4x \sqrt{-g}\mathcal{L}_{m}(g_{\mu\nu},\psi_m)$ is the matter action of a set of matter fields denoted collectively by $\psi_m$. Note that the independent connection $\Gamma$ does not enter in the construction of the matter sector, which will have relevant implications when discussing geodesic behaviour later.

The field equations of this theory are obtained by independent variation of the action (\ref{eq:actfR}) with respect to metric and connection, which yields the two systems of equations (see \cite{Olmo:2011uz} for a detailed derivation including torsion)
\begin{eqnarray}
&&f_{\mathcal{R}} R_{\mu\nu}-\frac{1}{2}g_{\mu\nu}f=\kappa^2 T_{\mu\nu} \label{eq:metricfR} \\
&&\nabla_{\beta}^{\Gamma}(\sqrt{-g}f_{\mathcal{R}}  g^{\mu\nu})=0 \label{eq:connectionfR}
\end{eqnarray}
where $f_{\mathcal{R}} \equiv \tfrac{df}{d\mathcal{R}}$ and $T_{\mu\nu} \equiv \tfrac{-2}{\sqrt{-g}} \tfrac{\delta S_m}{\delta g^{\mu\nu}}$ is the stress-energy tensor of the matter fields. It is thus immediately seen the deviance with respect to the metric formulation of $f(R)$ gravity (calligraphic letter removed to highlight its metric character): in such a case, the fact that $\Gamma$ is Levi-Civita of $g_{\mu\nu}$, replaces the two sets of equations above by the single equation
\begin{equation}
f_{R}R_{\mu\nu}-\frac{1}{2} g_{\mu\nu} f -[\nabla_{\mu}\nabla_{\nu} f_R-g_{\mu\nu} \Box R]=\kappa^2 T_{\mu\nu}
\end{equation}
The presence of two derivative operators acting upon the scalar function $f(R)$, which itself contains two derivatives, has two consequences: first the theory generally contains fourth-order equations of motion, and second, the object $\phi=f_R$ can effectively be seen as a new (scalar) propagating degree of freedom, which introduces some difficulties in order to make the theory compatible with weak-field limit observations \cite{Olmo:2005zr,Olmo:2005hc,Burrage:2017qrf}.  As opposed to this, in the Palatini formulation, such derivative operators are missing, which turns the field equations as second-order and there are no extra degrees of freedom to worry about. The price to be paid is that in this formulation one needs to solve the system of equations (\ref{eq:connectionfR}) in order to find the affine connection. In the present case, this is easily done by just noting that, after contraction of (\ref{eq:metricfR}) with the metric $g_{\mu\nu}$, one finds an algebraic equation $\mathcal{R}f_{\mathcal{R}}-2f(\mathcal{R})=\kappa^2T$, where $T=g^{\mu\nu}T_{\mu\nu}$ is the trace of the stress-energy tensor. This result tells us that $\mathcal{R}=\mathcal{R}(T)$, i.e., the curvature scalar can be written as a function of the trace $T$ of the matter fields. This implies that in the Palatini version of the theory, the scalar field carries no dynamics and its role will be reduced to introducing a deformation in the way the matter fields source the gravitational dynamics.

In turn, this result allows to solve the connection equations (\ref{eq:connectionfR}) by introducing a new rank-two tensor $q_{\mu\nu}$ related to the space-time metric and the matter fields via the (conformal) relation
\begin{equation} \label{eq:confrel}
q_{\mu\nu}=\frac{1}{f_{\mathcal{R}}(T)} g_{\mu\nu}
\end{equation}
where we recall that $f_{\mathcal{R}} \equiv df/d\mathcal{R}$ is a function of the trace $T$. In terms of $q_{\mu\nu}$,  Eq.(\ref{eq:connectionfR}) reads as $\nabla_{\beta}^{\Gamma}(\sqrt{-q}q^{\mu\nu})=0$, i.e., $\Gamma$ can be solved as the Christoffel symbols of $q_{\mu\nu}$. Equipped with the relation above, contracting again in (\ref{eq:metricfR}) with $g^{\alpha \mu}$, and suitably rearranging terms, one arrives at
\begin{equation} \label{eq:Rmunu}
{R^\mu}_{\nu}(q)=\frac{\kappa^2}{f_{\mathcal{R}}^2} \left(\frac{f}{2} \delta^{\mu}_{\nu} + \kappa^2 {T^\mu}_{\nu}\right)
\end{equation}
where ${T^\mu}_{\nu} \equiv T^{\mu\alpha} g_{\alpha\nu}$. Since $\Gamma$ is Levi-Civita of $q$, then ${R^\mu}_{\nu}(q)\equiv q^{\mu\alpha}R_{\alpha\nu}$ is the usual Ricci tensor computed with the Christoffel symbols of $q_{\mu\nu}$. Moreover, since both $f$ and $f_{\mathcal{R}}$ are functions of the matter sources, Eq.(\ref{eq:Rmunu}) is actually nothing more than Einstein equations (for $q$) coupled to a modified stress-energy tensor on its right-hand side. Therefore, they can be solved by resorting to the usual analytical and numerical methods developed within GR in order to find an expression for $q_{\mu\nu}$. Subsequently, using (\ref{eq:confrel}) it is trivial to find the corresponding solution for the space-time metric $g_{\mu\nu}$. The bottleneck of this procedure is to actually be able to solve the equation $\mathcal{R}=\mathcal{R}(T)$ under a workable enough form: for instance, in the quadratic case, $f(\mathcal{R})=\mathcal{R}+\alpha \mathcal{R}^2$, with $\alpha$ a constant with dimensions of length squared, one finds $\mathcal{R}=-\kappa^2 T$, which is actually the same result as in GR. For more involved functional dependencies, this would introduce additional difficulties in solving the field equations, so one would be forced to introduce numerical methods.


This simple extension of GR illustrates the benefits and drawbacks of working in the Palatini approach: on the former we find a system of second-order field equations without extra propagating degrees of freedom that can be solved with standard methods, while for the latter one needs to find first a suitable algebraic way to solve the connection equations feeding the metric one. This procedure can be generalized to other cases of interest, as we shall do next.

\subsection{Quadratic gravity}

The theory of quantized fields in a curved space-time tells us that an ultraviolet completion of GR must come under the form of higher-order contributions in the scalar objects out of curvature contractions, suppressed by inverse powers of the fundamental quantum (Planck) scale \cite{ParTom,Birrell}. This yields the natural generalization of $f(\mathcal{R})$ gravity by including another contraction of the Ricci tensor, $\mathcal{Q}=g^{\mu\alpha}g^{\nu\beta}R_{\mu\nu}R_{\alpha \beta}$, via an arbitrary function of these two invariants, i.e. $f(\mathcal{R},\mathcal{Q})$. Therefore, the theory (\ref{eq:actfRQ}) is generalized to
\begin{equation} \label{eq:actfRQ}
S=\frac{1}{2\kappa^2} \int d^4x \sqrt{-g} f(\mathcal{R},\mathcal{Q}) + S_m(g_{\mu\nu},\psi_m)
\end{equation}
with the same considerations and notations as before. The (Palatini) field equations associated to this action are again found by independent variations with respect to metric and connection, which in this case amount to
\begin{eqnarray}
&&f_{\mathcal{R}} R_{\mu\nu}-\frac{1}{2}g_{\mu\nu}f+2f_\mathcal{Q} R_{\mu\alpha}{R^\alpha}_{\nu}=\kappa^2 T_{\mu\nu} \label{eq:metricfRQ} \\
&&\nabla_{\beta}^{\Gamma}(\sqrt{-g}(f_{\mathcal{R}}  g^{\mu\nu}+2f_\mathcal{Q} R^{\mu\nu}))=0 \label{eq:connectionfRQ}
\end{eqnarray}
where a new contribution  in $f_\mathcal{Q} \equiv \tfrac{df}{d\mathcal{Q}}$ is apparent. In the attempt to implement the same strategy as in the $f(\mathcal{R})$ case to solve the connection equations, one first notes that by defining a tensor ${M^\mu}_{\nu} \equiv g^{\mu\alpha}R_{\alpha\nu}$, then the objects $\mathcal{R}$ and $\mathcal{Q}$ are simply given in terms of traces of this object as  $\mathcal{R}=\text{Tr}({M^\mu}_{\nu})$, $\mathcal{Q}=\text{Tr}({M^\mu}_{\alpha}{M^\alpha}_{\nu})$, respectively. The shape of each of these matrices is obtained from (\ref{eq:metricfRQ}) rewritten as  $2f_\mathcal{Q} \hat{M}^2+f_{\mathcal{R}}\hat{M}-\tfrac{f}{2}\hat{I}=\kappa^2 T$, where in all cases hats represent matrices. This implies that, similarly as in the $f(\mathcal{R})$ case above, one can solve this equation as $\hat{M} \equiv \hat{M}(\hat{T})$, i.e., both $\mathcal{R}$ and $\mathcal{Q}$ can be solved as functions of the stress-energy sources. In turn, this gives again consistency to introducing a new rank-two tensor as an attempt to solve the connection equations (\ref{eq:connectionfRQ}) via
$\sqrt{-q}q^{\mu\nu}=\sqrt{-g}g^{\mu\alpha}{\Sigma_\alpha}^{\nu}$, where the matricial object $\hat{\Sigma}=f_{\mathcal{R}}\hat{I}+2f_\mathcal{Q} \hat{M}$. Operating these relations in (\ref{eq:connectionfRQ}) one arrives at
\begin{equation} \label{eq:qtog}
q_{\mu\nu}= \vert \hat{\Sigma} \vert^{1/2} {\Sigma_\mu}^{\alpha}g_{\alpha\nu}
\end{equation}
where vertical bars denote a determinant. Furthermore, by contracting the metric field equations (\ref{eq:metricfRQ}) with ${M_\mu}^{\alpha}{\Sigma_\alpha}^{\nu}$ one arrives at
\begin{equation} \label{eq:eomfRQ}
{R_\mu}^{\nu}(q)=\frac{1}{\vert \hat{\Sigma} \vert^{1/2}} \left(\frac{f}{2}\delta_{\mu}^{\nu} + \kappa^2 {T_\mu}^{\nu} \right)
\end{equation} Since again $\Gamma$ is Levi-Civita of $q$, and both $\mathcal{R}$ and $\mathcal{Q}$ are functions of the matter sources, Eqs.(\ref{eq:eomfRQ}) can again be read off as a system of Einstein-like equations sourced with a modified stress-energy tensor on its right-hand side, now including those extra contributions from the $\mathcal{Q}$-piece. Similarly as in the $f(\mathcal{R})$ case, application of standard analytical/numerical methods to solve these equations for the metric $q_{\mu\nu}$, and subsequent use of the (matter-mediated) transformation (\ref{eq:qtog}) allows to put the corresponding results into a solution for the space-time metric $g_{\mu\nu}$. The bottleneck of this procedure is again to be able to find a workable solution for the matrix $\hat{\Sigma}$, which is more involved than in the $f(\mathcal{R})$ case given the disformal rather than conformal transformation (\ref{eq:qtog}) between $q$ and $g$. This resolution can be achieved for some simple enough examples, such as the quadratic gravity theory $f(\mathcal{R},\mathcal{Q})=\mathcal{R}+a\mathcal{R}^2 + b\mathcal{Q}$, with $(a,b)$ some parameters with dimensions of length squared.

It is worth pointing out that one could add to this quadratic gravity theory further contractions of the Riemann tensor such as the Kretchsmann scalar $K=R_{\alpha\beta\gamma\delta}R^{\alpha\beta\gamma\delta}$. However, we do not currently possess the necessary geometrical methods to deal with the resolution of the corresponding connection field equations. Instead, we shall head in the next section in a different direction.

\subsection{Eddington-inspired Born-Infeld gravity}

Adding further powers of the curvature scalars is not the only possible way to keep generalizing these theories. Indeed, every scalar Lagrangian of weight $-4$, appearing in the integral of the action, could make a candidate to a gravitational theory. A proposal whose relevance in the community has quite grown in the last few years was originally considered by Vollick \cite{Vollick:2003qp} and then popularized by Bañados and Ferreira \cite{Banados:2010ix}, being generally known as Eddington-inspired Born-Infeld gravity (EiBI), and is given by the action:
\begin{svgraybox}
{\bf Born-Infeld-type theories:} There is quite a long tradition of invoking square-root modifications of classical actions: from the relativistic Lagrangian of point particles or the Born-Infeld modification of classical electrodynamics to cure the electron's self-energy problem \cite{BI}, to the results of Fradkin and Tseytlin \cite{Fradkin:1985qd} showing that actions of the Born-Infeld type arise in different scenarios related to M-theory, to finally arrive to the gravitational arena. Indeed, Born-Infeld type formulations of the gravitational field have been considered according to several frameworks and to address many astrophysical phenomena (see \cite{BeltranJimenez:2017doy} for a review).
\end{svgraybox}
\begin{equation} \label{eq:actionEiBI}
S_{EiBI}=\frac{1}{\kappa^2 \epsilon} \int d^4x \left[ \sqrt{-\vert g_{\mu\nu} + \varepsilon R_{\mu\nu} \vert} - \lambda \sqrt{-g} \right] + S_m(g_{\mu\nu},\psi_m)
\end{equation}
where $\epsilon$ is a smallness scale with dimensions of length squared, encoding the deviations with respect to GR. In this sense, for $ \vert R_{\mu\nu} \vert \ll \epsilon^{-1}$, this theory boils down to
\begin{equation}
S_{EiBI} \approx \int d^4x \sqrt{-g} \left( \frac{\mathcal{R}}{2\kappa^2}-\Lambda_{eff}\right) -\frac{\epsilon}{4\kappa^2} \int d^4x \sqrt{-g} \left(-\frac{\mathcal{R}^2}{2}+\mathcal{Q} \right) + \mathcal{O}(\epsilon^2)
\end{equation}
which is nothing but GR with an effective cosmological constant $\Lambda_{eff}=\tfrac{\lambda-1}{\epsilon}$ supplemented with higher-order curvature corrections, which at the quadratic level it is actually an example of a $f(\mathcal{R},\mathcal{Q})$ theory. A Palatini formulation of this theory can be achieved if one introduces the new-rank two tensor as $q_{\mu\nu}=g_{\mu\nu}+\epsilon R_{\mu\nu}$, in such a way that a variation of the action (\ref{eq:actionEiBI}) with respect to the metric yields
\begin{equation} \label{eq:metEiBI}
\frac{\sqrt{-q}}{\sqrt{-g}} q^{\mu\nu} -g^{\mu\nu}= \kappa^2 \epsilon T^{\mu\nu}
\end{equation}
with (once again) $\Gamma$ being Levi-Civita of $q$. The above equations can be written in a more convenient (and workable) way by writing formally the relation between the two metrics of the theory under the algebraic relation
\begin{equation} \label{eq:funrel}
q_{\mu\nu}=g_{\mu\alpha}{\Omega^\alpha}_{\nu}
\end{equation}
To obtain the shape of the matrix $\hat{\Omega}$ we just need to contract the metric field equations (\ref{eq:metEiBI}) with the metric $g_{\nu\alpha}$ to find that its components  are given by the algebraic equation
\begin{equation} \label{eq:OmeEiBI}
\vert \Omega \vert^{1/2} {(\Omega^{-1})^\mu}_\nu=\lambda \delta^{\mu}_{\nu}-\kappa^2 \epsilon {T^\mu}_{\nu}
\end{equation}
which therefore can be determined once a matter sector is specified in the theory. Moreover, this matrix can also be used to rewrite the EiBI Lagrangian in (\ref{eq:actionEiBI}) in terms of it as
\begin{equation}
\mathcal{L}_G = \frac{\vert \Omega \vert^{1/2}}{\epsilon \kappa^2} - \lambda
\end{equation}
while suitably working upon the metric field equations (\ref{eq:actionEiBI}) one can rewrite them, with the help of $\hat{\Omega}$ and the expression above, as
\begin{equation}
{R^\mu}_{\nu}(q)=\frac{1}{\vert \Omega \vert^{1/2}}\left(\mathcal{L}_G \delta^{\mu}_{\nu} + \kappa^2 {T^\mu}_{\nu} \right)
\end{equation}
This way we arrive (once again) to a set of Einstein-like equations sourced with a modified stress-energy tensor. The above three examples manifest a general trend: by introducing a sort of ``auxiliary" metric $q_{\mu\nu}$ such that the independent connection is Levi-Civita of it, one can recast the field equations of the theory in Einstein-like form with its right-hand side emerging as the result of a sort of non-minimal coupling of the matter fields, while an algebraic {\it deformation} allows to transform the solution for $q_{\mu\nu}$ into a solution for the space-time metric $g_{\mu\nu}$. One can thus wonder what is the most general theory sharing all these features, which we tackle next.

\subsection{The Ricci-based gravity family}

We wish to build a (projectively-invariant) theory of gravity based on  every possible contractions of the (symmetric part of the) Ricci tensor with the metric. We can thus tentatively write
\begin{equation} \label{eq:RBGaction}
S_{RBG}=\frac{1}{2\kappa^2} \int d^4x \sqrt{-g}  \mathcal{L}_G(g_{\mu\nu},R_{(\mu\nu)})+S_m(g_{\mu\nu},\psi_m)
\end{equation}
In order to guarantee that the function $\mathcal{L}_G$ is an scalar object, its functional dependence on its argument must be via traces of the object ${M^\mu}_{\nu} \equiv g^{\mu\alpha}R_{(\alpha\nu)}$. These theories are dubbed as Ricci-based gravities (or RBGs for short) and includes, in particular, the three cases above, since $\mathcal{R}=\text{Tr}({M^\mu}_{\nu})$, $\mathcal{Q}=\text{Tr}({M^\mu}_{\alpha}{M^\alpha}_{\nu})$, while the EiBI Lagrangian comes from a combination of the polynomial invariants associated to ${M^\mu}_{\nu}$ (for a general description on how to build actions from such invariants see \cite{BeltranJimenez:2014hma}). To verify the validity of this general construction, we need to obtain the field equations associated to the  action (\ref{eq:RBGaction}). By performing independent variations with respect to the metric and the connection, and assuming an algebraic relation between them of the form (\ref{eq:funrel}), one arrives at the metric field equations (see \cite{Afonso:2017bxr} for a detailed derivation including torsion)
\begin{equation} \label{eq:RBGeom}
{G^\mu}_{\nu}(q)=\frac{1}{\vert \Omega \vert^{1/2}}\left({T^\mu}_{\nu}-\left(\mathcal{L}_G+\frac{T}{2}\right)\delta^{\mu}_{\nu} \right) \ .
\end{equation}
This equation confirms our expectations: that the whole RBG family admits an Einstein-like representation of its field equations (for the metric $q$) with its right-hand side being that of a modified stress-energy tensor, since $\mathcal{L}_G \equiv \mathcal{L}_G({T_\mu}^{\nu},g_{\mu\nu})$. The space-time metric $g_{\mu\nu}$ is obtained as an algebraic deformation of $q_{\mu\nu}$ via the fundamental relation (\ref{eq:funrel}) and, consequently, we shall dub $\hat{\Omega}$ as the {\it deformation matrix}. In physically sensible scenarios, this matrix will typically follow the same algebraic structure as the one of the stress-energy tensor ${T^\mu}_{\nu}$, though other solutions can also be found \cite{BeltranJimenez:2020guo}. Examples of this are the conformal transformation (\ref{eq:confrel}) of the $f(\mathcal{R})$ case, or the disformal transformations (\ref{eq:qtog}) and (\ref{eq:OmeEiBI}) of the $f(\mathcal{R},\mathcal{Q})$ and EiBI cases. Furthermore, the relations obtained so far imply that the deformation matrix is, in general, a function of both the matter fields and the space-time metric, thus the extra terms present in the right-hand side of the RBG field equations (\ref{eq:RBGeom}) will also be functions of the matter. These features have important consequences. Indeed, the RBG field equations are always (as opposed to their metric counterparts) second-order and, moreover, in absence of matter, ${T_\mu}^{\nu}=0$, its solutions (for $q$) are those of GR plus (possibly) a cosmological constant term, while as for the space-time metric itself one has $g_{\mu\nu}=q_{\mu\nu}$ (modulo a trivial re-scaling). This implies that RBG theories do not propagate other degrees of freedom beyond the two polarizations of the gravitational field (gravitational waves) travelling (in vacuum) at the speed of light (see \cite{Jana:2017ost} for the EiBI case), allowing them to naturally pass constraints from the observations of gravitational waves and their electromagnetic counterpart from binary neutron stars mergers \cite{GW1}. Modifications with respect to GR predictions occur only inside matter sources, where in addition to effects related to the total mass-energy of the sources  the new dynamics is also fed by energy-density effects, which is a feature absent within GR. This is yet another trademark of these theories with important consequences for the structure of their solutions.

There is, however, a difficulty to solve the field equations (\ref{eq:RBGeom}) due to the fact that, while their left-hand side is written in terms of $q_{\mu\nu}$, its right-hand side will be typically be written in terms of $g_{\mu\nu}$, whose inversion in terms of $q_{\mu\nu}$ using the fundamental relation (\ref{eq:funrel}) is not always easy (or even possible) depending of the algebraic properties of the deformation matrix in relation to the matter fields feeding it. In sufficiently symmetric scenarios, it is actually possible to invert this relation and write the field equations entirely in terms of $q_{\mu\nu}$, but for more complex (and interesting) scenarios, advanced techniques are required.

\subsection{Einstein frame and the mapping method}

The large resemblance of (\ref{eq:RBGeom}) with the canonical Einstein equations strongly suggests that a full Einstein representation should be possible, i.e.:
\begin{equation} \label{eq:eisteinthat}
{G^\mu}_{\nu}(q)=\kappa^2  \tilde{T}{^\mu}_{\nu}(q)
\end{equation}
for a new stress-energy tensor $\tilde{T}{^\mu}_{\nu}(q)$. To this end, let us introduce a set of auxiliary fields ${\Sigma^\mu}_{\nu}$ to rewrite the action (\ref{eq:RBGaction}) as
\begin{eqnarray}
S_{RBG}(g_{\mu\nu},\Gamma_{\mu\nu}^{\lambda},{\Sigma^\mu}_{\nu},\psi_m)&=&\frac{1}{2\kappa^2}\int d^4x \sqrt{-g}\left[\mathcal{L}_G({\Sigma^\mu}_{\nu}) +\left(g^{\mu\alpha}R_{\alpha\nu}-{\Sigma^\mu}_{\nu} \frac{\partial \mathcal{L}_G}{\partial {\Sigma^\mu}_{\nu}}\right)\right] \nonumber \\
&+&S_m(g_{\mu\nu},\psi_m)
\end{eqnarray}
It is easy to see that under a variation with respect to ${\Sigma^\mu}_{\nu}$ one finds that ${\Sigma^\mu}_{\nu}= g^{\mu\alpha}R_{\alpha\nu}$, provided that $\tfrac{\partial^2 F}{\partial {\Sigma^\mu}_{\nu}\partial {\Sigma^\rho}_{\lambda}} \neq 0$. Next, by introducing the auxiliary metric $q_{\mu\nu}$ via the definition
\begin{equation}
\sqrt{-q}q^{\mu\nu} \equiv \sqrt{-g}g^{\mu\alpha} \frac{\partial \mathcal{L}_G}{\partial {\Sigma^\alpha}_{\nu}}
\end{equation}
one finds that the action above can be rewritten as
\begin{equation}
S_{RBG}(g_{\mu\nu},\Gamma_{\mu\nu}^{\lambda},{\Sigma^\mu}_{\nu},\psi_m)=\frac{1}{2\kappa^2}\int d^4x \sqrt{-q}q^{\mu\nu}R_{\mu\nu}(\Gamma) +\tilde{S}_m(g_{\mu\nu},{\Sigma^\mu}_{\nu},\psi_m)
\end{equation}
where we identify this action as nothing but the Einstein-Hilbert action of GR (for $q$) with a non-minimally coupled matter field sector. Therefore, full consistence of this action with the Einstein-like field equations (\ref{eq:RBGeom}) has been achieved, but we are not nearer of solving our original riddle, namely, whether it is possible to write an stress-energy tensor satisfying the correspondence
\begin{equation} \label{eq:Tmunuhat}
 \tilde{T}{^\mu}_{\nu}(q) \equiv -\frac{2}{\sqrt{-q}} \frac{\delta \tilde{S}_m}{\delta q^{\mu\nu}}=\frac{1}{\vert \Omega \vert^{1/2}}\left({T^\mu}_{\nu}-\left(\mathcal{L}_G+\frac{T}{2}\right)\delta_{\mu}^{\nu} \right)
\end{equation}
removing systematically all the dependencies out in $g_{\mu\nu}$ on the right in terms of $q_{\mu\nu}$-contributions. As current research can tell us, this can only be done on a case-by-case basis of matter sources. To this end, let us consider quite a general scenario of anisotropic fluids given by the stress-energy tensor (in the RBG frame)
\begin{equation}\label{eq:Tmunufluid}
{T^\mu}_{\nu}=(\rho+P_{\perp})u^\mu u_\nu + (P_r-P_{\perp})\xi^{\mu} \xi_{\nu} + P_{\perp}\delta^{\mu}_{\nu}
\end{equation}
where we have introduced the unit (time-like and space-like vectors, respectively) $g_{\mu\nu}u^{\mu}u^{\nu}=-1$ and $g_{\mu\nu}\xi^\mu \xi^\nu=+1$, while $\rho$ is the energy density of the fluid, $P_r$ its pressure in the direction of $\xi^\mu$ and $P_{\perp}$ its pressure in the direction orthogonal to $\xi^\mu$. Note also that, in a comoving system, this stress-energy tensor can be simply cast as ${T^\mu}_{\nu}=\text{diag}(-\rho,P_r,P_{\perp},P_{\perp})$. Due to the orthogonality of these vectors, the deformation matrix must inherit the same algebraic structure, that is,
\begin{equation}
{\Omega^\mu}_{\nu}=\alpha \delta^{\mu}_{\nu}+\beta u^\mu u_\nu + \gamma \xi^\mu \xi_\nu
\end{equation}
where the expressions for the three functions $\{\alpha,\beta,\gamma\}$ are (both in RBG and matter sector) model-dependent. Inserting these expressions into the one for the stress-energy tensor in the GR frame, Eq.(\ref{eq:Tmunuhat}), yields
\begin{equation} \label{eq:Tsrel}
 \tilde{T}{^\mu}_{\nu}(q)=\frac{\kappa^2}{\vert \Omega \vert^{1/2}}\left[\left(\frac{\rho-P_r}{2}-F\right)\delta^{\mu}_{\nu}+(\rho+P_{\perp})u^{\mu}u_{\nu}+(P_r-P_{\perp})\xi^\mu \xi_\nu \right]
\end{equation}
so if we propose a formally similar expression as that of (\ref{eq:Tmunufluid}) but with new unit time-like $q_{\mu\nu}v^\mu v^\nu=-1$ and space-like $q_{\mu\nu}\chi^\mu\chi^\nu=+1$ vectors, and new functions characterizing the fluid, $\{\rho^q,P_r^q,P_{\perp}^q\}$, one finds the identifications
\begin{eqnarray}
P_{\perp}^q&=&\frac{1}{\vert \Omega \vert^{1/2}}\left[\frac{\rho-P_r}{2}-\mathcal{L}_G\right] \label{eq:map1} \\
\rho^q+P_r^q &=& \frac{\rho+P_r}{\vert \Omega \vert^{1/2}} \label{eq:map2}
\\
\rho^q-P_{\perp}^q &=& \frac{\rho-P_{\perp}}{\vert \Omega \vert^{1/2}} \label{eq:map3}
\end{eqnarray}
These relations allow to find a correspondence between the functions characterizing the fluid in the two frames once an RBG Lagrangian, given by a function $\mathcal{L}_G$, is specified. Furthermore, supplied with the relations $u^\mu u_\nu = v^\mu v_\nu$ and $\xi^\mu \xi_\nu =\chi^\mu \chi_\nu$, this also allows to write the deformation matrix in terms of those functions defining the fluid in the RBG frame. In turn, this allows one to solve the fundamental relation (\ref{eq:metEiBI}) in order to yield a solution for $g_{\mu\nu}$ once an expression for $q_{\mu\nu}$ is given. Since the latter corresponds to the solution of the GR problem coupled to a set of matter fields $\tilde{\psi}_m$ whose stress-energy tensor is given by $ \tilde{T}{^\mu}_{\nu}(q,\tilde{\psi}_m)$, the above relations {\it map} the seed solution $q_{\mu\nu}$ into a new solution $g_{\mu\nu}$ corresponding to a given RBG theory coupled to a set of matter fields $\psi_m$ with a stress-energy tensor ${T^\mu}_{\nu}(g,\psi_m)$. Moreover, working upon (\ref{eq:Tsrel}) this also allows (through a straightforward but tedious algebraic procedure) to find the correspondence between the gravitational Lagrangians on each side of the correspondence. It is worth pointing out that this result holds true irrespective of any symmetries involved in any problem under consideration, since we did not assume anything on the background solutions, but just worked at the level of the action and the general field equations.

The bottom line of this discussion is the power of the mapping method as a solution-generator machine. Indeed, it allows to generate solutions of a given pair $\{\text{RBG(g)},\psi_m\}$ starting from any known solution of the pair $\{\text{GR}(q),\tilde{\psi}_m \}$, with the correspondences above allowing one to find the relations between both the matter fields and the gravity Lagrangians. For the sake of this chapter we shall be interested in two particular cases of interest: scalar fields, which satisfy $P_{\perp}=P_{r}$, and (non-linear) electromagnetic fields, for which $P_r=-\rho$.

\begin{svgraybox}
{\bf The poltergeist nature of the mapping:} The mapping makes curious alchemy on the functional dependence of the action: for instance, quadratic $f(\mathcal{R})$ gravity coupled to a standard scalar/Maxwell field maps into GR coupled to quadratic scalar/Maxwell field, while EiBI gravity coupled to a Maxwell/Born-Infeld electrodynamics maps into GR coupled to a Born-Infeld/Maxwell electrodynamics!. This has side-effects in the way the structure of the corresponding solutions inherit properties from the original side of the mapping they came from.
\end{svgraybox}

This concludes our theoretical considerations on the theories and methods to be considered in this chapter. Next we shall head to the description of the most relevant black hole solutions found in the literature, before moving on to discuss their regularity.

\section{Spherically symmetric black hole solutions}

\subsection{A case-sample on the direct attack to solve the Palatini equations} \label{sec:1.3.1}

To start with our recollection of the black hole solutions of interest found in Palatini theories of gravity, let us first  analyze in detail a case-sample of how to solve their field equations via direct attack. This involves {\it a tale of two frames}, since we shall be solving the field equations for $q_{\mu\nu}$, i.e., Eqs.(\ref{eq:RBGeom}), while at the same time working out its relation with the space-time metric $g_{\mu\nu}$, i.e., Eq.(\ref{eq:funrel}). In order to simplify this discussion, let us consider the Palatini theory in which the relation between these two metrics is the simplest, as given by the conformal relation (\ref{eq:confrel}) of the $f(\mathcal{R})$ case. In this scenario, the fact that the new dynamics enters just via the trace $T$ of the stress-energy tensor prevents one from using any source of matter that yields a traceless stress-energy tensor, since the corresponding solutions would reduce to those of GR. This, unfortunately, leaves aside the interesting case of Maxwell electrodynamics, which would allow us to find the counterpart of the Reissner-Nordstr\"om black hole in these theories. Therefore, we shall consider non-linear electromagnetic fields, which are described by a function $\varphi(X,Y)$ of the field invariants
\begin{equation}
X=-\frac{1}{2}F_{\mu\nu}F^{\mu\nu} \ ;  \ Y=-\frac{1}{2}F_{\mu\nu}F^{\star\mu\nu}\ ,
\end{equation}
where $F_{\mu\nu}=\partial_{\mu}A_{\nu}-\partial_{\nu}A_{\mu}$ is the field strength tensor of the vector potential $A_{\mu}$ and $F^{\star\mu\nu}=\tfrac{1}{2}\epsilon^{\mu\nu\alpha\beta}F_{\alpha\beta}$ its dual. Since we are interested in static, spherically symmetric solutions, the magnetic part of the electrostatic field, built from the vector potential $A_{\mu}=(A_t,0,0,0)$, can be neglected, which entails $Y=0$. This way, the stress-energy associated to these fields, which reads
\begin{equation}
{T^\mu}_{\nu}=-\frac{1}{4\pi}\left(\varphi_X {F^\mu}_{\alpha}{F^\alpha}_{\nu}-\frac{\delta^{\mu}_{\nu}}{2}\varphi(X,0)\right) \ ,
\end{equation}
where $\varphi_X \equiv d\varphi/dX$, becomes in  the present case
\begin{equation}
{T^\mu}_{\nu}=\frac{1}{8\pi}\text{diag}(\varphi(X)-2X\varphi_X ,\varphi(X)-2X\varphi_X,\varphi(X),\varphi(X))
\end{equation}
On the other hand, we have the electromagnetic field equations
\begin{equation}
\nabla_\mu(\varphi_X F^{\mu\nu}+\varphi_Y F^{\star \mu\nu})=0 \ .
\end{equation}
Since these equations are coupled to the space-time metric $g_{\mu\nu}$, which in static, spherically symmetric configurations reads as
\begin{equation} \label{eq:sssmetric}
ds^2=g_{tt}dt^2+g_{rr}dr^2+r^2 d\Omega^2 \ ,
\end{equation}
where $d\Omega^2=d\theta^2+\sin^2 \theta d\phi^2$ is the volume element in the unit two-sphere, for electrostatic configurations these are solved as
\begin{equation}
\varphi_X F^{tr}=\frac{Q}{r^2\sqrt{-g_{tt}g_{rr}}} \ ,
\end{equation}
where $F^{tr}$ is the single non-vanishing component of the field strength tensor in this setting, while $Q$ is an integration constant identified as the electric charge for a given configuration. Moreover, using the fact that $X=-g_{tt}g_{rr}(F^{tr})^2$ we can rewrite the above equation as
\begin{equation} \label{eq:eomem}
X\varphi_X^2=\frac{Q^2}{r^4}
\end{equation}
which is an algebraic equation allowing to find the expression of the invariant $X$ (via a quadrature) when a NED Lagrangian is given. For further reference, we find it convenient to introduce at this stage the well known Born-Infeld (BI) electrodynamics, which reads
\begin{equation} \label{eq:BINED}
\varphi_{BI}(X,Y)=2\beta^2 \left(1-\sqrt{1-\frac{X}{\beta^2}+\frac{Y}{4\beta^2}} \right) \ ,
\end{equation}
where $\beta$ is the Born-Infeld parameter. For this theory, Eq.(\ref{eq:eomem}) tells us that $E(r)\equiv F^{tr}=\tfrac{\beta Q}{\sqrt{\beta^2 r^4+Q^2}}$. At large distances, $r \to \infty$, one finds that $E(r)$ reduces to the Maxwell field, $E(r) \approx Q/r^2$, while near the center, $r \to 0$, the finiteness of $E(r)$ makes the total energy associated to the electrostatic field to be finite. However, within GR this fact does not solve the singularity problem \cite{Diaz-Alonso:2009xkw}, since neither the completeness of geodesics nor the finiteness of curvature scalars is achieved. Further on this topic later.

Now that we have all the electromagnetic sector under control we can proceed with the resolution of the field equations (\ref{eq:Rmunu}). Given the symmetry of the stress-energy tensor in $2 \times 2$ blocks, such equations can be conveniently written as (here $\hat{I}_{2\times 2}$ and $\hat{0}_{2\times 2}$ are the identity and zero matrices, respectively)
\begin{equation} \label{eomem}
{R^\mu}_{\nu}(q)=\frac{1}{2f_R^2}
\begin{pmatrix}
(f+2\kappa^2 {T^t}_t) \hat{I}_{2\times 2} & \hat{0}_{2\times 2}  \\
\hat{0}_{2\times 2}  & (f+2\kappa^2 {T^\theta}_{\theta}) \hat{I}_{2\times 2}
\end{pmatrix}
\end{equation}
therefore retaining the symmetry of the matter source. To solve them we propose a static, spherically symmetric line element on the $q_{\mu\nu}$ frame as given by the convenient form
\begin{equation} \label{eq:dsq2}
ds_q^2=q_{\mu\nu}dx^\mu dx^\nu=-e^{2\psi(x)}A(x)dt^2 + \frac{dx^2}{A(x)}+x^2 d\Omega^2
\end{equation}
for the metric functions $\{\psi(x),A(x)\}$ parameterized in terms of a new radial coordinate $x$. To handle the left-hand side of the equations (\ref{eq:Rmunu}), one uses canonical methods to compute the components of the Ricci tensor (since it is built in terms of the Christoffel symbols of the line element (\ref{eq:dsq2})), with the result
\begin{eqnarray}
{R^t}_t&=& -\frac{A}{2}\left[\frac{A_{xx}}{A}-\left(\frac{A_x}{A}\right)^2+2\psi_{xx} +\left(\frac{A_x}{A}+2\psi_x \right) \left( \frac{A_x}{A} +\psi_x +\frac{2}{x} \right)  \right] \\
{R^r}_r&=&  -\frac{A}{2}\left[\frac{A_{xx}}{A}-\left(\frac{A_x}{A}\right)^2+2\psi_{xx} +\left(\frac{A_x}{A}+2\psi_x \right) \left( \frac{A_x}{A} +\psi_x \right) +\frac{2}{x}\frac{A_x}{A} \right] \\
{R^\theta}_{\theta}&=&(\sin^2 \theta) {R^\phi}_{\phi}=\frac{1}{x^2} \left[1-A(1+x\psi_x) -xA_x     \right] \label{eq:Rththcom}
\end{eqnarray}
Now, considering the subtraction ${R^t}_t-{R^r}_r=0$, the last equality being a trivial consequence of the $2\times 2$ block symmetry of the right-hand side of (\ref{eomem}), one finds that $\psi=$constant, which can be set to zero by a redefinition of the temporal coordinate without loss of generality. Plugging this result into (\ref{eq:Rththcom}), and working out the right-hand side of (\ref{eomem}), one finds that
\begin{equation}
\frac{1}{x^2}(1-A(x)-xA_x)=\frac{1}{2f_{\mathcal{R}}^2} \left(f+\frac{\kappa^2}{4\pi} \varphi \right)
\end{equation}
Next, proposing the usual mass ansatz
\begin{equation}
A(x)=1-\frac{2M(x)}{x}
\end{equation}
the above equation is solved as
\begin{equation} \label{eq:MxfR}
M_x \equiv \frac{dM}{dx}=\frac{x^2}{4f_{\mathcal{R}}^2}\left(f+\frac{\kappa^2}{4\pi} \varphi \right)
\end{equation}
To keep progressing we need to rewrite this last expression in terms of the variables of the space-time metric (\ref{eq:sssmetric}). This is easily done in view of the conformal transformation (\ref{eq:confrel}), which implies the relation between the radial coordinates in each frame as
\begin{equation} \label{eq:xrf}
x^2=r^2f_{\mathcal{R}}
\end{equation}

\begin{svgraybox}
{\bf The non-trivial radial function:} The tale of two frames usually ends up (at least in some branches of the solutions of the theory) in a non-trivial and non-monotonic behaviour of the radial function in the space-time metric in relation to the auxiliary one, i.e., $r^2(x)$. The consequences of this result (when it happens) for the regularity of the corresponding space-times must be discussed on a case-by-case basis, as shall be seen below.
\end{svgraybox}

Taking a derivative here one arrives at
\begin{equation}
\frac{dr}{dx}=\frac{1}{f_{\mathcal{R}}^{1/2}\left[1+\frac{rf_{\mathcal{R},r}}{2f_{\mathcal{R}}} \right]}
\end{equation}
and replacing it in (\ref{eq:MxfR}) we get
\begin{equation} \label{eq:Mr}
M_r \equiv \frac{dM}{dr} =\frac{r^2}{4f_{\mathcal{R}}^{3/2}}\left(f+\frac{\kappa^2}{4\pi} \varphi \right)\left(f_{\mathcal{R}}+rf_{\mathcal{R},r}\right)
\end{equation}
Typically, this function can be generically integrated as $M(r)=r_S(1+ \delta_1 G(r))/2$, where $r_S=2M_0$ is the Schwarzschild radius, $\delta_1$ encodes the relevant constants in the problem, and $G(r)$ is the primitive of $M_r$, thus a model-dependent function. Therefore, the space-time line element can be written as
\begin{equation} \label{eq:metricgfR}
ds_g^2=\frac{1}{f_{\mathcal{R}}}\left(-A(x)dt^2+\frac{dx^2}{A(x)}+r^2(x)d\Omega^2 \right)
\end{equation}
where the radial function is implicitly given by (\ref{eq:xrf}) while the metric function $A(r(x))$ is written as
\begin{equation} \label{eq:metrfR}
A(r)=1-\frac{1+\delta_1 G(r)}{\delta_2 rf_{\mathcal{R}}^{1/2}}
\end{equation}
where $\delta_2$ is another parameter collecting any additional constant that have emerged from our manipulations. This is how far we can go without further specifying our setting. Specific cases of interest can be now obtained by setting both a NED function $\varphi(X(r))$ and a shape for the $f(\mathcal{R})$ gravity function, which would allow to uniquely determine the right-hand side of (\ref{eq:Mr}) and, therefore, to provide a unique solution to this problem.

Let us thus set, for the gravity sector, the quadratic model
\begin{equation} \label{eq:quadraticgravf}
f(\mathcal{R})=\mathcal{R}-\sigma \mathcal{R}^2
\end{equation}
and for the matter sector the BI electrodynamics of Eq.(\ref{eq:BINED}). In order to work with dimensionless variables, it is convenient to introduce the following definitions: $r_q^2=\kappa^2 Q^2/4\pi$, $l_{\beta}^2=1/(\kappa^2 \beta^2)$ and the dimensionless radial function $z=r/r_c$, with $r_c^2=(4\pi)^{1/2}r_ql_{\beta}$. This way, Eqs.(\ref{eq:metrfR}) and (\ref{eq:Mr}) still hold [with the radial function $r(x)$ replaced by its dimensionless version $z(x)$, where an $r_c$ factor has also been reabsorbed in the $x$-coordinate], while the relevant gravity and matter functions become \cite{Olmo:2011ja}
\begin{equation}
\varphi(z)=2\left(1-\frac{z^4}{z^4+1} \right) \hspace{0.1cm} ; \hspace{0.1cm} f(z)=\frac{\eta(z)}{2\pi}\left(1-\frac{\alpha}{2} \eta(z)\right) \hspace{0.1cm}; \hspace{0.1cm} f_{\mathcal{R}}=1-\alpha \eta(z)
\end{equation}
where $\alpha \equiv \sigma/(2\pi l_{\beta}^2)$ and we have introduced the function
\begin{equation}
\eta(z)=\frac{(z^2-\sqrt{z^4+1})^2}{z^2\sqrt{z^4+1}}
\end{equation}
while the constants $\delta_1$ and $\delta_2$ appearing in (\ref{eq:metrfR}) read
\begin{equation}
\delta_1=2(4\pi)^{3/4} \frac{r_q}{r_S} \sqrt{\frac{r_q}{l_\beta}} \hspace{0.5cm} \ ; \ \hspace{0.5cm} \delta_2=\frac{r_c}{r_S}
\end{equation}
which completes our construction. The line element (\ref{eq:metricgfR}) with the definitions above represent a generalization of the Reissner-Nordstr\"om (RN) solution characterized by mass, charge, and gravity parameter. The RN solution is recovered in the limit $r \to \infty$, as can be verified by expansion of the metric functions in such a limit, i.e.,
\begin{equation}
-g_{tt} =g_{rr}^{-1} \approx 1-\frac{1}{\delta_2 z} + \frac{\delta_1}{16 \pi \delta_2 z^2} + \mathcal{O}(z^{-4})
\end{equation}
which after restoring back the $r$-notation is easily recognized as the actual Reissner-Nordstr\"om solution of GR, $-g_{tt}=g_{rr}^{-1}=1-\tfrac{2M}{r} + \tfrac{Q^2}{8\pi r^2} + \mathcal{O}(r^{-4})$. Deviations with respect to such a solution will occur as the new energy density contributions driven by the interplay between the quadratic $f(\mathcal{R})$ gravity and BI effects, and encoded in the single parameter $\alpha$, become relevant. This has important consequences for the global structure of the corresponding solutions in terms of the number and type of horizons, though for the sake of this part of the chapter we are more interested in describing the consequences for the regularity of these space-times, a topic that will be fully addressed in Sec. \ref{Sec:regularity}. In this sense, the expression of the radial function $z(x)$, obtained from Eq.(\ref{eq:xrf}) becomes also a critical aspect in the characterization of the regularity of these solutions, though an explicit expression is not always possible.

\subsection{Other spherically symmetric solutions}

Having concluded with our detailed analysis of the procedure to solve the field equations by direct attack, we shall continue elaborating our space of solutions, relevant for the discussion of regular black holes.  We shall first add other matter sources to our discussion of quadratic $f(\mathcal{R})$ gravity, and afterwards move to more complex gravitational theories. \\

{\it Quadratic $f(\mathcal{R})$ gravity with anisotropic fluids} \cite{Olmo:2015axa}. We next consider a version of the anisotropic fluid (\ref{eq:Tmunufluid}) given by
\begin{equation}
{T^\mu}_{\nu}=\text{diag}(-\rho,-\rho,\alpha \rho,\alpha \rho)
\end{equation}
where the constant $\alpha$ is constrained within the range $0<\alpha \leq 1$ to satisfy the classical energy conditions, with the upper limit $\alpha=1$ corresponding to the usual Maxwell electrodynamics. We consider the coupling of this matter source to the quadratic $f(\mathcal{R})$ model (\ref{eq:quadraticgravf}). The energy density of the fluid can be found via integration of the conservation equation of the stress-energy tensor (this conservation being a consequence of the fact that the independent connection does not enter into the matter sector), $\nabla_{\mu}T^{\mu\nu}=0$, as $\rho=\rho_0/r(x)^{2(1+\alpha)}$, where $\rho_0$ is a (dimensionful) integration constant. The resolution of this problem amounts to the line element (\ref{eq:metricgfR}), where the new dimensionless scale is defined as $r_c^{2(1+\alpha)}=(4\sigma) \kappa^2 \rho_0 (1-\alpha)$ so that the constant $\delta_1=r_c^3/(4\sigma r_S)$ and the relevant function characterizing the metric reads (for $\alpha \neq 1/2$):

\begin{svgraybox}
{\bf Classical energy conditions:} Classical matter fields must satisfy several conditions to fulfil fundamental facts on our understanding about the geometrical and physical components of our Universe. These conditions are typically given by the following \cite{Martin-Moruno:2017exc} (we do not consider here averaged energy conditions):
\begin{itemize}
\item Null energy condition (NEC): $T_{\mu\nu}n^\mu n^\nu \geq 0$, with $n^\mu n_\mu =0$ a null vector. NEC implies $\rho+P_i \geq 0$ $\forall i=1,2,3$.
\item Weak energy condition (WEC): $T_{\mu\nu}u^\mu u^\nu \geq 0$, with $u^\nu u_\nu =-1$ a time-like vector. WEC implies $\rho \geq 0$ and $\rho+P_i \geq 0$ $\forall i=1,2,3$.
\item Strong energy condition (SEC): $T_{\mu\nu}u^\mu u^\nu \geq -T/2$. SEC implies $\rho +\sum_{i=1}^3 P_i \geq 0$ and $\rho + P_i \geq 0$ $\forall i=1,2,3$.
\item Dominant energy condition (DEC): $-{T^\mu}_{\nu} u^\mu$ is a future-oriented null or time-like vector. DEC implies $\rho \geq 0$ and $\rho \geq \vert P_i \vert$  $\forall i=1,2,3$.
\end{itemize}
The reach of such conditions is inherently limited by the fundamental quantum nature of the matter fields; for instance, the Casimir effect violates them \cite{Roman:1986tp}.
\end{svgraybox}
\begin{equation}\label{eq:G}
G(z)=\frac{z^{-4 \alpha -1} \left(\frac{z^{2 \alpha +2} \sqrt{1-z^{-2 (\alpha +1)}} \left(2 \alpha ^2+\alpha +2 z^{2 \alpha +2}-3\right)}{z^{2 \alpha +2}-1}-\frac{8 \alpha  \left(\alpha ^2-1\right) \, _2F_1\left(\frac{1}{2},\frac{4 \alpha +1}{2 \alpha +2};\frac{6 \alpha +3}{2 \alpha +2};z^{-2 (\alpha +1)}\right)}{4 \alpha +1}\right)}{2 (\alpha -1) (2 \alpha -1)} \ ,
\end{equation}
where $_2F_1\left(a, b, c; y\right)$ is a hypergeometric function. As for the behaviour of the radial function in these solutions, it is implicitly given by the expression
\begin{equation}
x=\frac{\sqrt{z^{2(1+\alpha)}-1}}{z^\alpha}
\end{equation}
Likewise in the previous case, this solution can be seen as a deformation of the RN solution of GR inside the matter sources, modifying the structure of horizons and having relevant implications for the regularity of these space-times. \\

{\it Quadratic $f(\mathcal{R})$ gravity with more general anisotropic fluids} \cite{Bejarano:2017fgz}: Let us consider now the more general anisotropic fluid
\begin{equation} \label{eq:anisoflu}
{T^\mu}_{\nu}= \text{diag}(-\rho,-\rho,K(\rho),K(\rho))
\end{equation}
where the free function $K(\rho)$ characterizes the fluid, with $K(\rho)=\alpha \rho$ for the previous case. NEDs are naturally included within this class of fluids via the identifications $-8\pi \rho=\varphi -2(X\varphi_X + Y\varphi_Y),8\pi K(\rho)=\varphi$. In order to find explicit solutions, let us consider the ansatz for the free function as $K(\rho)=\rho+\beta \rho^2$  and introduce the definitions $\tilde{\beta}=s_{\beta} \vert \beta \vert \rho_0$ (with $s_{\beta}=\pm 1$ the sign of $\beta$) and $r_c=(\beta \rho_0)^{1/4}r_0$. Then the density of the fluid becomes $\rho(z)=\tfrac{\rho_m}{z^4-s_{\beta}}$, with $\rho_m=2/\vert \beta \vert$, and one also finds the constant $\delta_1=\kappa^2\rho_m(r_0 \vert \tilde{\beta} \vert^{1/4})^3/r_S$.  Now there are four different solutions for the function $G_z$ depending on the signs of $\beta$ and $\gamma$ as (here we have defined $\gamma \equiv 8\kappa^2 \rho_m \vert \sigma \vert$)
\begin{eqnarray}
G_z&=& \frac{z^2 \left(1-\frac{\gamma  \left(1-3 z^4\right)}{\left(z^4+1\right)^3}\right) \left(1-\frac{\gamma }{\left(z^4+1\right)^3}\right)}{\left(z^4+1\right) \left(1-\frac{\gamma }{\left(z^4+1\right)^2}\right)^{3/2}} \hspace{0.2cm} \text{for $\sigma>0, \beta<0$} \label{eq:Gfluid1} \\
Gz&=& \frac{z^2 \left(1-\frac{\gamma  \left(3 z^4+1\right)}{\left(z^4-1\right)^3}\right) \left(1-\frac{\gamma }{\left(z^4-1\right)^3}\right)}{\left(z^4-1\right) \left(\frac{\gamma }{\left(z^4-1\right)^2}+1\right)^{3/2}} \ \hspace{0.2cm} \text{for $\sigma>0, \beta>0$} \label{eq:Gfluid2} \\
G_z&=& \frac{z^2 \left(1+\frac{\gamma  \left(1-3 z^4\right)}{\left(z^4+1\right)^3}\right) \left(1+\frac{\gamma }{\left(z^4+1\right)^3}\right)}{\left(z^4+1\right) \left(1+\frac{\gamma }{\left(z^4+1\right)^2}\right)^{3/2}}  \hspace{0.2cm} \text{for $\sigma<0, \beta<0$} \label{eq:Gfluid3} \\
G_z&=& \frac{z^2 \left(1+\frac{\gamma  \left(1+3 z^4\right)}{\left(z^4+1\right)^3}\right) \left(1+\frac{\gamma }{\left(z^4-1\right)^3}\right)}{\left(z^4-1\right) \left(1-\frac{\gamma }{\left(z^4-1\right)^2}\right)^{3/2}} \hspace{0.2cm} \text{for $\sigma<0, \beta>0$} \label{eq:Gfluid4} \ .
\end{eqnarray}
and that we shall call them Type I, Type II, Type III, and Type IV, respectively.  Accordingly, the behaviour of the radial function $z(x)$ is different in each case. In particular, only when $\sigma$ and $\beta$ have opposite signs, do the radial function $z(x)$ implements a bouncing behaviour. \\

{\it Quadratic gravity with Maxwell electrodynamics} \cite{Olmo:2011np,Olmo:2012nx,Lobo:2013adx}. Let us now upgrade our setting to consider the case of the quadratic model
\begin{equation}
f(\mathcal{R},\mathcal{Q})=\mathcal{R}+l^2(a\mathcal{R}^2+\mathcal{Q})
\end{equation}
where $l$ is a new parameter with dimensions of length  characterizing the scale of the corrections, and  $a$ is a dimensionless constant. As the matter source, the fact that the new gravitational dynamics associated to these theories has access to the full structure of the ${T^\mu}_{\nu}$, and not just to its trace, like in the $f(\mathcal{R})$ case, allows us to consider a standard Maxwell field as the matter source. Therefore, we set $\varphi(X)=X$ and proceed to solve the corresponding field equations. To this end, a straightforward but tedious algebraic exercise acting upon the trace of the metric field equations (\ref{eq:metricfRQ}), allows one to find $\mathcal{Q}=\tfrac{\tilde{\kappa}^2 Q^2}{r^8}$, with $\tilde{\kappa}^2 \equiv \kappa^2/4\pi$. It is illustrative now to write the full expression of the RBG field equations (\ref{eomem}) in this case as
\begin{equation}
{R^\mu}_\nu(q)=\frac{\tilde{\kappa}^2Q^2}{2r^4}\begin{pmatrix}
-\frac{1}{\Omega_+} \hat{I}& \hat{0} \\
\hat{0} & \frac{1}{\Omega_-} \hat{I}
\end{pmatrix}  \label{eq:Ricci-h4} \ .
\end{equation}
where  we have defined the objects $\Omega_{\pm}=1-1/z^4$ and in this case $r_c^4=\kappa^2 Q^2 l^2$. Note that the conformal factor $f_{\mathcal{R}}$ of the previous cases degenerates into the objects $\Omega_{\pm}$, which manifests itself at several levels in the structure of the corresponding solution. To find the latter, one follows the same tale of the two frames as in the quadratic $f_{\mathcal{R}}$ case above, which yields the result
\begin{eqnarray}
ds^2&=&-\frac{A(z)}{\Omega_+}dt^2 +\frac{1}{\Omega_+ A(z)}dx^2+z^2(x)d\Omega^2 \hspace{0.1cm} ; \hspace{0.1cm} A(z)=1-\frac{1+\delta_1 G(z)}{\delta_2 z \Omega_{-}^{1/2}} \label{eq:quadraticgraveq1} \\
\delta_1&=&\frac{1}{2r_S} \sqrt{\frac{r_q^3}{l}};  \delta_2=\frac{r_c}{r_S} \hspace{0.1cm}; \hspace{0.1cm} G_z=\frac{\Omega_+}{z^2 \Omega_{-}^{1/2}} \label{eq:quadraticgraveq2}
\end{eqnarray}
where in this case the radial function satisfies $x^2=z^2\Omega_{-}$, which can be inverted to find the explicit expression
\begin{equation} \label{eq:quadraticgraveq3}
z^2(x)=\frac{x^2+\sqrt{x^2+4}}{2}
\end{equation}
We have yet another generalization of the RN solution of GR. However, this also introduces interesting modifications to the global structure of the solutions, in the sense that if $\delta_1>\delta_c$, with $\delta_c \approx -0.572$, one finds the presence of two (Cauchy and event) horizons, likewise in the RN solution of GR, while if $\delta_1<\delta_c$ then a single horizon is present, thus resembling more the Schwarzschild black hole of GR. The case $\delta_1 = \delta_c$ has peculiar properties beyond its horizon structure (having a single horizon or not, depending on the absolute value of the electric charge), which will be very relevant in our discussion of the regularity of black hole solutions. We point out that similar comments on the horizons and critical configurations apply to solutions out of fluids discussed previously in this section.  \\

{\it EiBI gravity with Maxwell electrodynamics} \cite{Olmo:2013gqa}. The EiBI gravity given by the action (\ref{eq:actionEiBI}) is particularly amicable for this sort of computations. This is so thanks to the simpler structure of the equation for the deformation matrix, Eq.(\ref{eq:OmeEiBI}). Indeed, once the stress-energy tensor of the matter fields is specified (in particular, its symmetry in blocks), finding a solution to (\ref{eq:OmeEiBI}) is as simple a mathematical exercise as in their quadratic gravity cousins. Indeed, for a Maxwell field, $\varphi(X)=X$, the solution to this equation reads simply
\begin{equation} \label{eq:sigma+-}
{\Omega^{\mu}}_{\nu} =
\begin{pmatrix}
 \Omega^{(\epsilon)}_{+} \hat{I} & \hat{0}  \\
\hat{0} & \sigma^{(\epsilon)}_{-} \hat{I}
\end{pmatrix} \ , \ \Omega^{(\epsilon)}_{\pm}=\lambda \pm \frac{\vert s_\epsilon \vert}{z^4}
\end{equation}
where now $r_c^4=r_q^2 l_{\epsilon}^2$ with the new scale $\epsilon=-2l_{\epsilon}^2$, and $s_\epsilon= \pm 1$ is the sign of $\epsilon$. The field equations in this case become
\begin{equation}\label{eq:Rmn}
{R^{\mu}}_{\nu}(q)=\frac{1}{\epsilon}\left(
\begin{array}{cc}
\frac{(\Omega^{(\epsilon)}_{-} -1)}{\Omega^{(\epsilon)}_{-} }\hat{I}& \hat{0}  \\
\hat{0}& \frac{(\Omega^{(\epsilon)}_{+} -1)}{\Omega^{(\epsilon)}_{+} }\hat{I} \end{array}
\right)
\end{equation}
At this state one can note that, in all the solutions studied so far, the deformation matrix has a similar algebraic structure in blocks as the stress-energy tensor, and this feature is inherited by the corresponding field equations. To solve them, one follows the same strategy as in the previous case, with the surprising result that one gets exactly to the same equations (\ref{eq:quadraticgraveq1}), (\ref{eq:quadraticgraveq2}) and (\ref{eq:quadraticgraveq3}) as in the quadratic case. This coincidence is not accidental, as shall be clear in the discussion of the mapped solutions below. \\

{\it $f(\mathcal{R})$ gravity and EiBI gravity with Euler-Heisenberg electrodynamics} \cite{Guerrero:2020uhn}. In order to compare quadratic $f(\mathcal{R})$ and EiBI gravity solutions on equal footing, we can consider, for instance, the coupling of both of them to the same NED, which we take to be given by the Euler-Heisenberg (EH) electrodynamics. The action of the latter is written as (again, we restrict ourselves to purely electrostatic solutions, so $Y=0$)
\begin{equation}
\varphi(X)=X+\beta X^2
\end{equation}
The corresponding equations of motion can be solved via a quadrature, with the resulting expression for the field invariant \cite{Kruglov:2017ymn}:
\begin{equation}\label{Eq:X_red}
X(z)=\dfrac{1}{6 \beta\, z^{4/3}}\left[ \left( 1+\sqrt{1+z^4}\right)^{1/3}+\left( 1-\sqrt{1+z^4}\right)^{1/3} \right]^2.
\end{equation}
where $r_c^4=54\pi r_q^2 l_\beta^2$ and $l_{\beta}^2=\beta/\kappa^2$. Coupling this field to the quadratic $f(\mathcal{R})$ gravity model (\ref{eq:quadraticgravf}) yields the solution (\ref{eq:metricgfR}) and (\ref{eq:metrfR}) with the following definitions
\begin{eqnarray}
f &=& \dfrac{2}{9 \pi} \tau^4 (z) \left(1-\dfrac{\tilde{\sigma}}{2}\tau^4(z)\right) \hspace{0.1cm}; \hspace{0.1cm} f_R = 1 - \tilde{\alpha}\,  \tau^4(z) \\
	\varphi &=& \frac{\tau^2(z)}{6\pi}\left(1+\frac{2}{3}\tau(z)\right) \hspace{0.1cm};\hspace{0.1cm}
	\tau(z) = \text{Sinh} \left[\frac{1}{3} \ln \left[\frac{1}{z^2}\left(1+ \sqrt{z^4+1}\right)\right] \right] \label{eq:tau} \\
\delta_1&=& \frac{(54\pi)^{3/4}}{2r_S} \sqrt{\tfrac{r_q^3}{l_{\beta}}}
\end{eqnarray}
and $\tilde{\sigma} \equiv 4\sigma/(9\pi l_{\beta}^2)$. A bounce in the radial function is present in the branch $\tilde{\sigma}>0$ and its minimum is located at $z_c=\sqrt{2a/(a^2-1)}$ with $a=\exp[3 \text{ArcSinh}(\vert \tilde{\sigma}\vert^{-1/4}]$.

Now, considering the same problem now with EiBI gravity (\ref{eq:actionEiBI}), one arrives to the line element (\ref{eq:quadraticgraveq1}), where the relevant functions are now defined as (here $l_{\epsilon}^2=\epsilon/(12\pi l_{\beta}^2)$)
\begin{eqnarray} \label{eq:GzBI}
G_z &=& z^2 (\Omega_--1)\; \Omega_-^{1/2}\left(1+ \dfrac{z  \Omega_{-},{z}}{2 \, \Omega_-} \;\right) \label{Eq:G(z)_EiBI}   \\
\Omega_{+} &=& \lambda -l_{\epsilon}^2 \,\tau^2(z) \left(1+\frac{ 2\,\tau^2(z)}{3} \right) \hspace{0.1cm};  \hspace{0.1cm}
\Omega_{-} = \lambda + l_{\epsilon}^2\,\tau^2(z) \,(1+2\,\tau^2(z)) \ , \label{eq:Om-bi} \\
\delta_1 &=& \dfrac{r_c^3}{r_S \, \epsilon}  \hspace{0.1cm}  ;   \hspace{0.1cm}
 \delta_2 = \dfrac{r_c}{r_S} \ ,
\end{eqnarray}
In this case, only for $l_{\epsilon}^2>0$ does the radial function $z(x)$ attain a minimum at a certain $z_0$. \\

{\it EiBI gravity with anisotropic fluids} \cite{Menchon:2017qed}. The anisotropic fluid configuration (\ref{eq:anisoflu}) with $K(\rho)=\rho+\beta \rho^2$ solved in the quadratic $f(\mathcal{R})$ case above also admits an exact solution in the EiBI gravity case. The solution is again given by the line element (\ref{eq:quadraticgraveq1}) and similar definitions as in the $f(\mathcal{R})$ case while the relevant functions characterizing the metric are now given by the compact expressions
\begin{equation}
G_z= \frac{z^2 \Omega_1}{ (z^4-s_{\beta})\Omega_2^{1/2}} \hspace{0.1cm}  ; \hspace{0.1cm} \Omega_1=1-s_{\epsilon} \xi^2 \left( \frac{z^4+s_{\beta}}{(z^4-s_{\beta})^2} \right) \hspace{0.1cm}; \hspace{0.1cm}
\Omega_2=1+\frac{s_{\epsilon} \xi^2}{z^4-s_{\beta}} \label{eq:Gz}
\end{equation}
where now $\delta_1= \frac{  r_{c}^3}{r_S l_m^2}$, $l_m=\beta/(2\kappa^2)$ and  $\xi=l_{\epsilon}^2/l_{\beta}^2$ . Likewise in the quadratic $f(\mathcal{R})$ case, the corresponding solutions split into four different types depending on the combinations of the signs of $s_{\beta}=\beta /\vert \beta \vert$ and $s_{\epsilon}=\epsilon/\vert \epsilon \vert$. In Type-I $\{s_{\epsilon}=-1,s_{\beta}=-1\}$ a bounce is present at $z_c=(\xi^2-1)^{1/4}$ provided that $\xi^2>1$, in Type-II $\{s_{\epsilon}=-1,s_{\beta}=+1\}$ a bounce is always present at $z_c=(1+\xi^2)^{1/4}$, while in Type-III $\{s_{\epsilon}=+1,s_{\beta}=-1\}$ and in Type-IV $\{s_{\epsilon}=+1,s_{\beta}=+1\}$ no bounces are present.   \\

{\it Functional extensions of EiBI gravity with Maxwell electrodynamics} \cite{Bambi:2016xme}: By regarding the object $\vert \Omega \vert$ as the main building block in constructing a metric-affine action, one can consider a family of functional extensions of EiBI gravity defined by the action
\begin{equation}
S=\frac{1}{\kappa^2 \epsilon} \int d^4x \sqrt{-g}[ f(\vert \Omega \vert)-\lambda] + S_m(g_{\mu\nu},\psi_m)
\end{equation}
where the choice $f(\vert \Omega \vert)=\vert \Omega \vert^{1/2}$ corresponds to the usual EiBI gravity Lagrangian. Similar methods as in the other members of the RBG family can be applied in this case and, in particular, the tale of two frames still holds. Using such methods, and considering again a Maxwell field as the matter source, one can obtain an exact solution for the family $f(\vert \Omega \vert)=\vert \Omega \vert^{n/2}$  under (once again) the form (\ref{eq:quadraticgraveq1}), now with the definitions
\begin{equation}
G_z=-z^2\frac{(\vert \hat{\Omega} \vert^{\frac{n}{2}}-\sigma_{+})\Omega_{-}^{1/2}}{n^2 \vert \hat{\Omega} \vert^{\frac{2n-1}{2}}}  \Bigg[ (n-1)\vert \hat{\Omega} \vert^{\frac{n}{2}} \Bigg(1+\frac{n}{4} \frac{z \vert \hat{\Omega} \vert_z}{\vert \hat{\Omega} \vert} \Bigg)+\sigma_{+} \Bigg] \ . \label{eq:diffg}
\end{equation}
where now we have to upgrade our definitions of the $\Omega_{\pm}$ matrices to
\begin{equation} \label{eq:omega-p}
\Omega_{\pm}= (n-1)\vert \hat{\Omega} \vert^{\frac{n}{2}} + \sigma_{\pm} \hspace{0.1cm}; \hspace{0.1cm}
\sigma_{\pm}= \lambda \mp X \ .
\end{equation}
and similar definitions for the $z$ variable and constant $\delta_1$ as in the quadratic gravity and EiBI cases apply. Again, for $\epsilon<0$ and in the cases $1/2<n\leq 1$ a bounce in the coordinate $z(x)$ driven by the relation $x=z\Omega_{-}^{1/2}$ is found.  \\

{\it EiBI gravity with scalar fields} \cite{Afonso:2017aci}. A scalar field with Lagrangian density $\mathcal{L}_m=X-V(\phi)$, where in this case $X=\partial_\mu \phi \partial^\mu \phi$, can be seen, in the static, spherically symmetric setting, $ds_g^2=-A(x)dt^2+B^{-1}(x)dx^2+r^2(x)d\Omega^2$, as a sort of anisotropic fluid with stress-energy tensor
\begin{equation}
{T^\mu}_{\nu}=\text{diag}\left(-\frac{L_m}{2},B\phi_x^2-\frac{L_m}{2},B\phi_x^2-\frac{L_m}{2},B\phi_x^2-\frac{L_m}{2}\right)
\end{equation}
where $\phi_x \equiv d\phi/dx$. Therefore we see that the main novelty of this case as compared to the electromagnetic and fluid ones studied before is that the $2\times 2$-block structure of the stress-energy of the former is replaced by a $1\times 3$-block one in the latter. However, the methods of the previous cases and the tell of two frames work equally fine (though with some technical adjustments), and one proposes two suitable line elements of the form (here $C_0$ is a constant)
\begin{eqnarray}
d{s}_g^2&=&-e^{{\nu}}dt^2+\frac{1}{C_0^2 {W}^4 e^{-\nu}}dx^2+\frac{1}{{W}^2}d\Omega^2 \ ,\label{ds}\\
ds_q^2&=&-e^{\tilde{\nu}}dt^2+\frac{1}{C_0^2 \tilde{W}^4 e^{-\tilde{\nu}}}dy^2+\frac{1}{\tilde{W}^2}d\Omega^2 \  ,\label{dst}
\end{eqnarray}
subject to the relations $e^{\tilde{\nu}}=\Omega_+e^{\nu};
\tilde{W}^2=W^2/\Omega_+; dy= \frac{\Omega_-}{\vert \Omega \vert^{1/2}} dx= |\lambda-X_\epsilon|^{-1} dx$, where $X_\epsilon= \tfrac{\epsilon \kappa^2 C^2}{2\,r^4A}$ and the functions $\Omega_{+}=A_+^{1/2}A_{-}^{1/2},\Omega_{-}=A_+^{3/2}A_{-}^{1/2}$ with $A_{\pm}=(\lambda + \epsilon \kappa^2 V \pm \tfrac{\epsilon \kappa^2}{2} B\phi_x^2)$. The corresponding field equations can be conveniently written (in the $q$-frame) as
\begin{eqnarray}
\tilde{\nu}_{yy}&=&-\frac{\kappa_0^2 \Omega_+^3}{X_\epsilon}\left(1-\tfrac{\lambda+X_\epsilon}{\Omega^{1/2}}\right)  \label{fulleq1}\\
0&=& \tilde{W}_{yy}-\tilde{\nu}_y\tilde{W}_y- \frac{ \kappa_0^2\Omega_+^3}{2\Omega^{1/2}}\tilde{W} \label{fulleq2}\\
\tilde{W}\tilde{W}_{yy}-\tilde{W^2_y}&=& -\frac{e^{\tilde{\nu}}}{C_0^2}+ \frac{\kappa_0^2 \Omega_+^3}{2X_\epsilon} \left(1-\tfrac{\lambda+X_\epsilon}{\Omega^{1/2}}\right)\tilde{W}^2  \  ,   \label{fulleq3}
\end{eqnarray}
where $\kappa_0^2\equiv \kappa^2 v_0^2$ ($v_0$ another integration constant) has dimensions of length$^{-2}$, while $C_0^2$ has dimensions of length$^4$. These equations are impermeable to analytical resolution, so one has to resort to numerical methods. The latter reveal the presence of solitonic-like configurations in the spectrum of solutions, with non-trivial consequences for the analysis of the regularity of these space-times. \\

{\it Higher-dimensional solutions}. For the sake of completeness of this section, let us mention that higher-dimensional solutions can be found for the RBG family using the flexibility of the framework developed above. Indeed, the RBG field equations (\ref{eq:RBGeom}) are naturally generalized to this case as
\begin{equation}
{G^\mu}_{\nu}(q)=\frac{\kappa^2}{\vert \Omega \vert^{\frac{1}{(d-2)}}}\left({T^\mu}_{\nu}-\left(\mathcal{L}_G+\frac{T}{2}\right)\delta_{\mu}^{\nu} \right)
\end{equation}
where $d$ is the number of space-time dimensions and $\mathcal{L}$ the gravity Lagrangian. Assuming a Maxwell field, now with $X=Q^2/(r^{2(d-2)}$, and the EiBI action generalized to higher-dimensions, remarkably the line element (\ref{eq:quadraticgraveq1}) still holds [with the extensions of the volume element part of the unit sphere to the $(D-2)$-dimensional space-time], while the relevant functions become now \cite{Bazeia:2015uia}
\begin{eqnarray}
A(z)&=&1-\left(\frac{1+\delta_1 G(z)}{\delta_2  \Omega_{-}^{\frac{d-3}{2}}  z^{d-3}  }\right) \hspace{0.1cm} ; \hspace{0.1cm} G_z =-z^{d-2} \left( \frac{\Omega_{-}-1}{\Omega_{-}^{1/2}} \right) \left( \lambda + \frac{1}{z^{2(d-2)}} \right) \\
\Omega_{-}&=&\left(\lambda + \frac{1}{z^{2(d-2)}}\right)^{\frac{2}{d-2}}  \hspace{0.1cm} \hspace{0.1cm}; \hspace{0.1cm} \Omega_{+}=\frac{\left(\lambda - \frac{1}{z^{2(d-2)}} \right)}{\left(\lambda + \frac{1}{z^{2(d-2)}}\right)^{\frac{d-4}{d-2}}}
\end{eqnarray}
where now the relevant constants read
\begin{equation}
\delta_1 \equiv \frac{(d-3)r_c^{d-1}}{r_S l_{\epsilon}^2} \hspace{0.1cm}; \hspace{0.1cm} \delta_2 \equiv \frac{ (d-3)r_c^{d-3}}{r_S}.
\end{equation}
with $\epsilon=-l_{\epsilon}^2$ and $r_c^{2(d-2)} \equiv l_{\epsilon}^2\kappa^2 Q^2/(4\pi)$. Bouncing behaviours in the radial function $z(x)$ are again found in the branch $\epsilon <0$. Similarly, solutions can be found for certain $f(\mathcal{R})$ gravity models
 \cite{Bazeia:2014xxa}; for instance, in the model $f(\mathcal{R})=\mathcal{R} - \sigma \mathcal{R}^{5/2}$ the corresponding trace equations (recall that in higher dimensions the trace of the stress-energy tensor of a Maxwell field is non-vanishing) yields a simple solution of the curvature scalar, $\mathcal{R}=-\tfrac{2r_q^4}{3r^6}$, which in turn allows to find analytical solutions. Indeed, for every $f(\mathcal{R})$ theory, bouncing behaviours will be found according to the zeroes of the equation $x^2=z^2f_{\mathcal{R}}^{(2-n)/2}$, which will be present in the branch $\sigma >0$.

\subsection{Mapping-generated solutions}

Having discussed a bunch of solutions obtained from direct attack of the field equations, we now turn to those generated via the mapping. Now, the focus of our discussion comes to working out the mapping equations (\ref{eq:map1}), (\ref{eq:map2}) and (\ref{eq:map3}), finding the correspondences between the stress-energy tensors on both GR/RBG sides, and starting from a seed solution of the former reconstructing the corresponding action plus matter field on the latter. Let us discuss some examples.

\begin{svgraybox}
{\bf Mastering the mapping:} The mapping is a powerful but difficult-to-tame tool. It generically maps the same kind of matter fields into each other, though described by different Lagrangians and coupled to different RBGs. After the identification of such a combination of gravity + matter is worked out (which does {\it not} depend on any particular symmetry of the problem under consideration), its application to finding new exact solutions from a known {\it seed} solution (the latter typically within GR) is outrageously simple. See \cite{Afonso:2018bpv} for a conceptual discussion of this procedure.
\end{svgraybox}

{\it Mapping electromagnetic fields} \cite{Afonso:2018mxn}.   As mentioned above, (non-linear) electromagnetic fields can be seen as anisotropic fluids, which  we parameterize, in the RBG frame, via a function $\varphi(X)$ with stress-energy tensor
\begin{equation}
{T^\mu}_{\nu}=\text{diag}(-\rho,-\rho,K_{\varphi}(\rho),K_{\varphi}(\rho))
\end{equation}
and in the GR frame by a $\Phi(Z)$ with
\begin{equation}
\tilde{T}{^\mu}_{\nu}=\text{diag}(-\tilde{\rho},-\tilde{\rho},\tilde{K}_{\Phi}(\rho),\tilde{K}_{\Phi}(\rho))
\end{equation}
The mapping is particularly helpful when relating GR and EiBI gravity, where the resulting expressions are remarkably simple, being given by (tildes indicate an implicit factor $\epsilon \kappa^2$)
\begin{equation}
\tilde{\rho}_{\text{\tiny EiBI}}=\frac{\lambda\tilde{\rho}_{\text{\tiny GR}}-(\lambda-1)}{1-\tilde{\rho}_{\text{\tiny GR}}} \hspace{0.1cm}; \hspace{0.1cm}
\tilde{K}_{\text{\tiny EiBI}}=\frac{\lambda \tilde{K}_{\text{\tiny GR}}+(\lambda-1)}{1+\tilde{K}_{\text{\tiny GR}}} \label{eq:fluidNEDs}
\end{equation}
where the labels ``GR" and ``EiBI" refer to the respective sides of the correspondence. Taking $\lambda=1$, one can reconstruct the corresponding Lagrangians on each side as
\begin{equation}
\text{EiBI} + X \ \leftrightarrow \ \text{GR} + \hat{\Phi}(Z)=\frac{1}{2}\left(-1+\sqrt{1+\frac{\epsilon \kappa^2 Z}{2\pi}}\right) \label{EXtoGR}
\end{equation}
On the left-hand side of this correspondence we have the same setting of EiBI gravity with a Maxwell field discussed in the previous section, a problem that the reader may remember we solved via direct attack of the field equations, while on the right-hand (GR) side we recognize the BI electrodynamics upon the identification $\beta^2=-2\pi/(\epsilon \kappa^2)$, which only holds true in the branch $\epsilon <0$. The latter is a well known problem that admits a general solution for every NED as
\begin{equation}
M(x)=\frac{r_S}{2}+\frac{\kappa^2}{2} \int_r^{\infty} x^2 {T^t}_t(x)dx
\end{equation}
where ${T^t}_{t}$ is the stress-energy tensor of the (in the present case) BI electrodynamics with the identification made above. This expression feeds the mapping equations (\ref{eq:fluidNEDs}) together with the fluid's conservation equation discussed in previous section, which amounts to the equation $\rho(1-\epsilon \kappa^2 \rho) =\tfrac{Q^2}{8\pi x^4}$, and which upon integration yields exactly the expression (\ref{eq:quadraticgraveq2}) of the quadratic gravity/EiBI gravity case.

The main lesson of this quick exercise is to show how hardly-won solutions can be found via a much direct procedure. Note that the correspondence of theories works in both ways, i.e., if in the mapping equation (\ref{eq:fluidNEDs}) we set Maxwell in the GR side and assume EiBI as the target RBG theory on the other side of the correspondence, then one can reconstruct the corresponding Lagrangians as
\begin{equation}
\text{EiBI} + \Phi(X)=\frac{4\pi}{\kappa^2 \epsilon} \left(1-\sqrt{1-\frac{\kappa^2 \epsilon X}{2\pi}} \right) \leftrightarrow \text{GR} + Z \label{eq:EtoGR}
\end{equation}
where in the EiBI side of the correspondence we see again the appearance of a BI-type electrodynamics, with the right signs of the constants under the integral if we assume the identification $\beta^2=2\pi/(\kappa^2 \epsilon)$, where one could decide to impose $\epsilon >0$ in order to match the expectation of $\beta^2$ being a positive quantity in the GR scenario (although one could also explore the branch with $\epsilon<0$ oblivious to where this solution came originally from). Via this identification one can use once again the fluid's mapping  equations (\ref{eq:fluidNEDs}) to arrive to the line element  (\ref{eq:quadraticgraveq2}), in this case with the mass function
\begin{equation}
M(z)=\frac{r_S}{2}-\frac{\kappa^2 Q^2}{8\pi \sqrt{2} r_c}\frac{1}{\sqrt{z^2+\sqrt{z^4+4s}}}
\end{equation}
where now $r_c=\vert \epsilon \vert\kappa^2 Q^2/(8\pi)$ and $s\equiv \epsilon/\vert \epsilon \vert$ is the sign of $\epsilon$. As for the radial function, it behaves in this case as $z^2(x)=\tfrac{x^4-s}{x^2}$, which implies that a bounce is present in the branch $s=-1$ only.

Since the structure of the mapping above works for any electromagnetic field configuration, there is no limit to what we can achieve now in these scenarios starting from as many known seed solutions as desired. For instance, the Majumdar-Papapetrou configurations \cite{Maj,Pap}  are a subset of solutions of the Einstein-Maxwell-dust system in which the mass is
exactly tuned to the electric charge, and which can be interpreted as a collection of extreme black holes in static equilibrium given by the line element \cite{HarHaw}
\begin{equation} \label{eq:MP1}
ds^2=-U^{-2}(R)dt^2 +U^2 (dR^2 +R^2 (d\theta^2 + \sin^2 \theta d\phi^2)) \ ,
\end{equation}
where the function $U(r)$ satisfies the equation $\nabla^2  U=-4\pi \rho \,U^3$ thus granting some freedom to choose either the energy density $\rho$ or the $U(r)$ field itself. For instance, the so-called Bonnor stars \cite{LemZan} are described by an external solution $U(r)=1+\tfrac{m}{r}$ and an internal one $U(r)=1+\tfrac{m}{r_0}+\tfrac{m(r_0^2-r^2)}{2r_0^3}$ for a certain $r \geq r_0$. This scenario corresponds to the identification (\ref{eq:EtoGR}) under the map, so working again the relation between fluids (\ref{eq:fluidNEDs}) in the present case, we get to the general solution of the corresponding problem in the EiBI+ BI side, which in the spherically symmetric case reads \cite{Olmo:2020fnk}
\begin{equation}
ds^2=-\frac{1}{U^2}\left(1+\frac{\epsilon \kappa^2}{16\pi} \frac{(\nabla U)^2}{U^4}\right) dt^2 +\frac{\epsilon \kappa^2}{8\pi} \frac{(dU)^2}{U^2}
+  U^2 \left(1-\frac{\epsilon \kappa^2}{16\pi} \frac{(\nabla U)^2}{U^4}\right) d\vec{x}^2 \ ,
\end{equation}
where the EiBI+BI corrections in $\epsilon$ to GR solutions are evident. This actually describes a family of wormholes in static equilibrium, with also relevant consequences for the regularity of the corresponding space-times.

For the sake of completeness, it is worth pointing out that the mapping also works in higher and lower-dimensional versions of the RBG theories, which therefore allows one to consider seed solutions of interest from GR within such cases. For instance, the BTZ solution \cite{BTZ} describes a family of rotating configurations in the $2+1$ dimensional Einstein-Maxwell systems with cosmological constant, which interpolates between black holes and a pure AdS space-time, including the case of a regular horizonless solution with mass $M=-1$ disconnected from the space of black hole solutions by a mass gap. Since the correspondence (\ref{eq:EtoGR}) works equally fine in this case, a solution to the equivalent problem in the EiBI+BI side of the correspondence is found as \cite{Guerrero:2021avm}
\begin{eqnarray} \label{eq:BTZ_EiBI_EM}
\frac{ds^2}{\eta} &=&\left(M+\frac{Q^2}{2}\log \left(\frac{x}{x_0}\right)-\frac{x^2}{l^2} -\frac{\epsilon \kappa^2 J^2 Q^2}{64\pi \eta x^4}\right)dt^2 +H^{-1}dx^2 \\
&+&x^2\left(1-\frac{\epsilon \kappa^2 Q^2}{16\pi \eta x^2}\right)d\phi^2-J\left(1-\frac{\epsilon \kappa^2 Q^2}{16\pi \eta x^2}\right)dtd\phi \ , \nonumber
\end{eqnarray}
where $x_0$ is some reference scale,  $\eta\equiv (1-2\epsilon \Lambda)$, $H(x)=-M-\frac{Q^2}{2}\log (x/x_0)+\frac{x^2}{l^2}+\frac{J^2}{4x^2}$, and we have taken $\kappa^2=8\pi$ and $\Lambda=-1/l^2$. As usual, the sign of $\epsilon$ controls whether in the new $\epsilon$-deformed configurations there is ($s_{\epsilon}=-1$) or not ($s_{\epsilon}=+1$) a bounce in the radial function. \\

{\it Mapping scalar fields}. Similar correspondences to those of (\ref{EXtoGR}) and  (\ref{eq:EtoGR}) also hold when one considers scalar fields as the matter source, though now the algebraic structure of the corresponding stress-energy tensor makes quite more difficult to prove it so  \cite{Orazi:2020mhb}. Specifically, assuming a scalar Lagrangian in the RBG frame given by $\mathcal{L}_m=-\tfrac{1}{2}P(X,\phi)$, with $X=g^{\mu\nu}\partial_\mu \phi \partial_\mu \phi$, and another scalar Lagrangian in the GR frame as $\tilde{\mathcal{L}}_m=-\tfrac{1}{2}K(Z,\phi)$, with $Z=q^{\mu\nu}\partial_\mu \phi \partial_\nu \phi$, one can show the explicit shape of such correspondences as (in the quadratic $f(\mathcal{R})$ and EiBI cases)  \cite{Afonso:2018hyj}
\begin{eqnarray}
f(\mathcal{R})&=&\mathcal{R}-\sigma \mathcal{R}^2 + (P=X-2V) \leftrightarrow \text{GR} +K=\frac{Z+\sigma\kappa^2Z^2}{1+8\sigma\kappa^2 V}-\frac{2V}{1+8\sigma\kappa^2 V} \label{eq:mappscalar1} \\
f(\mathcal{R})&=&\mathcal{R}-\sigma \mathcal{R}^2 + P=\frac{X-\sigma\kappa^2X^2}{1-8\sigma\kappa^2V}-\frac{2V}{1-8\sigma\kappa^2V}  \leftrightarrow \text{GR} + K=Z-2V \label{eq:mappscalar2} \\
\text{GR}&+& K=Z-2V \leftrightarrow \text{EiBI} + P=\frac{2 \left(\sqrt{1+ \epsilon \kappa ^2 X}-\lambda(1+\epsilon \kappa ^2 V)\right)}{\epsilon \kappa ^2 \left(1+\epsilon \kappa ^2 V)\right)} \label{eq:mappscalar3}
\end{eqnarray}
Let us recall that the power of the mapping reaches its apex when a seed solution is known under analytical exact form in the GR side. In the present case, this is achieved for the free scalar case, where the corresponding solution was found by Wyman in \cite{Wyman}. Therefore, setting $V=0$ in Eqs.(\ref{eq:mappscalar2}) and (\ref{eq:mappscalar3}) one gets a quadratic-like scalar field Lagrangian in the quadratic $f(\mathcal{R})$ gravity case, and a square-root Born-Infeld-type in the EiBI case, yet another manifestation of the transfer of the functional dependencies between the gravity and matter Lagrangians when moving from one side of the mapping to the other. With these ingredients, finding the corresponding Wyman-like solutions in the quadratic $f(\mathcal{R})$ and EiBI frames is immediate and read \cite{Afonso:2019fzv}
\begin{equation} \label{eq:f(R)le}
ds_{f(R)}^2=\frac{1}{f_R} ds^2_{GR}=(1+2\alpha \kappa^2Z)\Big[-e^{\nu}dt^2+\frac{e^\nu}{W^4}dy^2 + \frac{1}{W^2}(d\theta^2+\sin\theta^2d\varphi^2) \Big] \ ,
\end{equation}
where $Z \equiv W^4 e^{-\nu}$ the field invariant of the Wyman case, and
\begin{equation}
ds_{EiBI}^2=-e^{\nu}dt^2+\left(\frac{e^\nu}{W^4}-\epsilon \kappa^2\right)dy^2 +\frac{1}{W^2}(d\theta^2+\sin\theta^2d\varphi^2)  \ , \label{eq:BImapping}
\end{equation}
respectively. \\

This concludes our presentation of the space of solutions found within several Palatini theories of gravity, obtained either via direct attack of the field equations or via the shortcut provided by the mapping method. It is now time to turn to the analysis of the regularity of all these configurations.

\section{Regularity criteria} \label{Sec:regularity}

\subsection{Curvature divergences}

Since (most) modern gravitational theories (including GR itself) see gravitation as a manifestation of a geometrical effect, then when a ``singularity" is present in a given configuration our physical intuition tells us that this must be caused by something going ill with the underlying geometrical structure of the space-time \cite{Curiel}. In order to avoid considering artificial singularities caused out of a bad choice of basis on our tensorial quantities, in characterizing singularities it seems natural to resort to objects that are invariant under coordinate transformations, i.e., scalar geometrical objects. Natural among them are curvature scalars, namely, different contractions of the Riemann tensor with the metric. This intuitive view amounts to build objects like $g_{\mu\nu}R^{\mu\nu},R_{\mu\nu}R^{\mu\nu},{R^\alpha}_{\beta\gamma\delta}{R_\alpha}^{\beta\gamma\delta},\ldots$, and consider a space-time as singular whenever {\it any} of such scalar objects is divergent. For instance, Schwarzschild space-time would be singular because in this case $g_{\mu\nu}R^{\mu\nu}=R_{\mu\nu}R^{\mu\nu}=0$ but ${R^\alpha}_{\beta\gamma\delta}{R_\alpha}^{\beta\gamma\delta} =48M^2/r^6$, and therefore the geometry is singular at $r=0$ due to the divergence of the latter invariant there. {\it Mutatis mutandis}, a space-time has a chance of being regular if all possible curvature scalars one can think of are finite. This procedure is so popular, besides its physical intuitiveness, thanks to its easy implementation: it can be actually programmed in any computation software such as Mathematica with a few lines of code (at least in the spherically symmetric case), and applied to any configuration of mathematical/physical interest.

Since most of our solutions are deformations of the Reissner-Nordstr\"om solution of GR inside the matter sources, it is instructive to begin our analysis by considering the behaviour of its curvature scalars. The three more ``popular" ones are given by
\begin{equation}
R_{RN}=0 \ , \ Q_{RN}=\frac{r_q^4}{r^8} \ , \ K_{RN}=\frac{12 r_S^2}{r^6}-\frac{24 r_S r_q^2}{r^7}+\frac{14 r_q^4}{r^8} \ .
\end{equation}
\begin{svgraybox}
{\bf Projective invariance and curvature scalars:} The fact that neither the Ricci nor the Riemann tensor are invariant under projective transformations, makes the scalar object $K$ to be non-projectively invariant as well ($\mathcal{Q}$ is projectively invariant since it is built from the symmetric part of the Ricci tensor alone, which is invariant), i.e., $K(\tilde{\Gamma})=K(\Gamma)+4F_{\mu\nu}F^{\mu\nu}$. This means that one can always find a suitable gauge for $\xi_\mu$ compensating the divergences in $K$ so as to render it finite everywhere \cite{Bejarano:2019zco}. However, there is yet another scalar object, $P\equiv {R^{\alpha\beta}}_{\mu\nu}{R_{\alpha\beta}}^{\mu\nu}$, which is invariant both under coordinate and projective transformations. This invariant typically diverges as strongly as $K$ itself, so that such divergences cannot be so simply gauged away. For the sake of this section we shall take the gauge $\xi_\mu =0$, where the scalar invariants take their canonical expressions.
\end{svgraybox}

How can one then find regular space-times via the behaviour of curvature scalars? Instead of writing off different matter sectors and solve their corresponding field equations to see whether they give rise to regular solutions or not, one can use the backward procedure: to {\it engineer}  space-times with any desired form of its line element, and then drive the Einstein equations back in order to find the matter sources threading such a geometry. In particular, one could seek space-times having a chosen set of their curvature scalars finite\footnote{Note, however, that one should establish a criterion to define a fundamental set of scalar invariants because the number of possible scalar functions is infinite.}. Focusing on the spherically symmetric case, and using the fact that the conflictive location of such scalars corresponds typically to $r=0$, this analysis finds that in order to acquire the finiteness of such invariants, the metric must behave there as \cite{Dymnikova:2003vt}
\begin{equation}
-g_{tt}=g_{rr}^{-1} \approx 1 - \frac{\Lambda}{3} r^2 + \mathcal{O}(r^3)
\end{equation}
where $\Lambda$ is a constant.  This is such a strong departure from the behaviour of the RN metric there, $-g_{tt} \sim +Q^2/r^2$, that one may wonder what type of matter fields could generate it. Actually, NEDs are able to do the job \cite{Dymnikova:2004zc}, though the corresponding models do not come free of theoretical difficulties as every other model does in trying to overcome the constraints of the singularity theorems (at least within GR). This kind of configurations go collectively under the name of {\it de Sitter} cores \cite{Ansoldi:2008jw,Lemos:2011dq}, and are a well known mechanism invoked in the literature for the removal of space-time singularities.

Is it possible to find curvature singularity-free solutions within Palatini theories of gravity without resorting to engineering constructions or de Sitter cores. To illustrate this issue, we consider our already familiar quadratic gravity/EiBI geometry of Eq.(\ref{eq:quadraticgraveq1}). For convenience we introduce Eddington-Finkelstein coordinates $dv=dt+\tfrac{dx}{A}$, which turns such a line element into
\begin{equation}
ds^2=-\frac{A}{\Omega_{+}}dv^2+\frac{2dvdx}{\Omega_+} +z^2(x)d\Omega^2
\end{equation}
Since the object $\Omega_{+}$ is positive and well-behaved everywhere, it cannot be a source of singularities anywhere. Troubles can thus only come out of the behaviour of the function $A(x)$, similarly as in GR. Therefore, expanding the function $B(x)=A(x)/\Omega_+$ in the line element above in series around the minimum value of the radial function $z=1$ ($r=r_c$ in dimensionful coordinates), one finds
\begin{eqnarray}
B(z)&\approx&  \frac{(\delta_1-\delta_c)}{4\delta_c\delta _2}\left(\frac{1}{ \sqrt{z-1}}+\frac{9}{4} \sqrt{z-1}-\ldots\right)+\nonumber \\ &+&\frac{1}{2}\left(1-\frac{\delta _1}{ \delta _2}\right)+\left(1-\frac{2\delta _1}{3\delta _2}\right) (z-1)+\mathcal{O}(z-1)^2 \label{eq:gtt_series}
\end{eqnarray}
where we recall that $\delta_c$ is a constant coming out of the resolution of the field equations. Therefore we see that in this case the leading-order divergence is quite mild, $\sim (z-1)^{-1/2}$, and significantly improves the RN behaviour. Moreover, this expressions also hints that the class of solutions with $\delta_1=\delta_c$ may have even improved properties. Therefore, replacing this condition first in the general expression of the $B(z)$ function, and expanding the result again in power series, we get the result
\begin{equation}
B(z)\approx \frac{1}{2}\left(1-\frac{\delta_c}{\delta _2 }\right)+\left(1-\frac{2\delta_c}{3 \delta _2 }\right)(z-1)-\frac{1}{2}\left(1-\frac{8\delta_c}{5 \delta _2 }\right)(z-1)^2+\mathcal{O}(z-1)^3   ,
\end{equation}
which is finite as $z \to 1$ and singles out this case as a special class of configurations.  Note that these expressions also allow to classify the structure of horizons of the corresponding solutions, which can be two (Cauchy and event), a single one (degenerate or not) or none, depending on the combination of parameters. Therefore, we may have either black holes with different horizon structures, or horizonless compact objects.

With the expressions above for $B(z)$ one can now compute the behaviour of the corresponding curvature scalars. In the general case, these read
\begin{eqnarray}
r_c^2 R(g)&\approx &\left(-4+\frac{16 \delta_c}{3 \delta _2}\right)+O\left({{z-1}}\right)+\ldots  \\ &-& \frac{1}{2 \delta _2 }\left(1-\frac{\delta_c}{\delta _1}\right)\left[\frac{1}{(z-1)^{3/2}}-O\left(\frac{1}{\sqrt{z-1}}\right)\right]  \nonumber, \\
r_c^4 Q(g) &\approx &  \left(10+\frac{86 \delta _1^2}{9 \delta _2^2}-\frac{52 \delta _1}{3 \delta _2}\right)+O\left({{z-1}}\right)+\ldots  \\
&+&\left(1-\frac{\delta_c}{\delta _1}\right)\left[\frac{6 \delta _2-5\delta _1}{3 \delta _2^2 (z-1)^{3/2}}+O\left(\frac{1}{\sqrt{z-1}}\right)\right] \nonumber \\
&+&\left(1-\frac{\delta_c}{\delta _1}\right)^2\left[\frac{1}{8 \delta _2^2 (z-1)^3}-O\left(\frac{1}{({z-1})^2}\right)\right] \nonumber \\
r_c^4K(g)&\approx & \left(16+\frac{88 \delta _1^2}{9 \delta _2^2}-\frac{64 \delta _1}{3 \delta _2}\right)+O\left({{z-1}}\right)+\ldots  \\
&+&\left(1-\frac{\delta_c}{\delta _1}\right)\left[\frac{2 \left(2\delta _1-3 \delta _2\right)}{3 \delta _2^2 (z-1)^{3/2}}+O\left(\frac{1}{\sqrt{z-1}}\right)\right]+\nonumber \\ &+&\left(1-\frac{\delta_c}{\delta _1}\right){}^2\left[\frac{1}{4 \delta _2^2 (z-1)^3}+O\left(\frac{1}{({z-1})^2}\right)\right] \nonumber.
\end{eqnarray}
From these expressions we observe that curvature divergences are softened from $\sim 1/r^8$ of RN to $\sim 1/(z-1)^3$ in the present case. Moreover, in the special case $\delta_1=\delta_c$, these scalars become all finite and given by the constant terms appearing in this expressions [note that again one arrives to this result by first substituting $\delta_1=\delta_c$ in the expression of $B(z)$ and then finding the corresponding shapes of them]. This points out to the striking nature of these $\delta_1=\delta_c$ configurations: they correspond to regular (in the sense of free of curvature divergences) black holes with a single non-degenerated horizon if $\delta_1>\delta_2$, and to horizonless regular objects otherwise, thus covering both scenarios of interest.

Two questions come about naturally. Does this mean that those configurations with $\delta_1=\delta_c$ are free of singularities, and those with $\delta_1 \neq 0$ are ``singular"?. Is this a general property of all configurations in Palatini theories of gravity?. The answer to the latter is negative, as is proven by the  case of quadratic $f(\mathcal{R})$ gravity coupled either to a Born-Infeld \cite{Olmo:2011ja} or to Euler-Heisenberg electrodynamics \cite{Guerrero:2020uhn}, where the invariant $K$ is divergent at the center of the solutions no matter the choice of their parameters. However, other matter configurations show different features. Indeed, for curvature scalars to be finite, we already saw before that the metric at the center must be finite, and this can be achieved for certain fluid configurations and in certain combination of parameters of the gravity+matter models. This is the case, for instance, of the configurations of Eq.(\ref{eq:Gfluid2}). Here there is no bounce in the radial function (as opposed to the case of a quadratic/EiBI gravity with a Maxwell field discussed above), but the energy density becomes divergent at $z=1$. Because of this feature, it is natural to see the latter surface as the boundary of the space-time, and consider the expansion of the metric functions there, which yields the result
\begin{equation}
-g_{tt} \approx \frac{r_S \delta_1}{3r_c\gamma}+\mathcal{O}(z-1) \hspace{0.1cm};\hspace{0.1cm} g_{rr} \approx \frac{48r_c}{r_S \delta_1} +\mathcal{O}(z-1)^2
\end{equation}
As a consequence of this latter behaviour, all curvature scalars for these configurations are also finite. One could of course argue that the divergence of the energy density at $z=1$ in these configurations would be a strong argument against the regularity and/or physical plausibility of these space-times. There are yet more examples of this kind within the setting above though now in a different branch of the theory as defined by Eq.(\ref{eq:Gfluid3}), where one finds that while there is no bounce in the radial function, in this case the configurations can reach the center $z=0$, where the metric functions become
\begin{equation}
-g_{tt} \approx \frac{1}{1+\gamma}\left(1-\frac{r_S \delta_c^\gamma}{3r_c} z^2  \right) \hspace{0.1cm}; \hspace{0.1cm} g_{rr} \approx 1+\frac{r_S \delta_c^\gamma}{3r_c}z^2
\end{equation}
where $\delta_c^\gamma$ is a constant playing the same role as its namesake in the quadratic/EiBI gravity case discussed above. In such a case, the geometry becomes de Sitter at the center, $R_{\mu\nu}=\tfrac{\delta_c^\gamma}{\delta_2}g_{\mu\nu}$, which yields finite curvature scalars with finite energy density everywhere, thus removing the objections above on the plausibility of these configurations. Indeed we note that, as opposed to GR-based attempts, this curvature-regularity via de Sitter cores can be naturally achieved with physically sensible matter fields satisfying basic energy conditions.

It turns out to be a generic property of $f(\mathcal{R})$ theories that solutions implementing a bounce in the radial function are prone to develop curvature divergences at the bounce location no matter the choice of the (gravity+matter) model parameters, while those with a monotonic behaviour of the radial function contain some sub-cases where the presence of curvature divergences can be removed everywhere. On the opposite side, we have quadratic/EiBI gravity, in which most configurations have curvature divergences at the bounce (whenever it exists), but at the same time some special configurations (as given by the choice of model parameters) can be found for which curvature scalars are finite everywhere.

However, the characterization of the degree of singularity of a space-time via divergences in the curvature scalars as pathology markers has been long contested in the literature. Arguments against such a line of reasoning come in several types. Firstly, given the many geometrical scalars that can be built beyond the $\{R,Q,K\}$ trio, in this approach one should be systematic in computing all possible such objects from every possible geometrical entity in order to make sure that they are all finite. Second, it could happen that every possible such scalar is regular and still pathological behaviours are present. But even more importantly,  one may question this privileged role attributed to the behaviour of some geometrical quantities rather than to the behaviour of observers instead, driven by the implicit idea that an ill-behaviour of the former will necessarily be translated on something pathological occurring to the latter. Indeed, this view on the prominence of curvature scalars can be discredited by presenting explicit counter-examples in which the blow-up of (some) curvature scalars do not necessarily translate into any pathologies of observers propagating through the space-times holding them \cite{Bejarano:2019zco}.  This idea connects with a more powerful concept in order to pin down the potential presence of pathologies in a given space-time, namely, the completeness of geodesic paths on them.



\subsection{Geodesic completeness and mechanisms for its restoration}

Placing the focus upon the idea of geodesic completeness over that of curvature divergences is simply a recognition of the motto that {\it the very existence of observers is more important that their potential suffering}. From a physical point of view, the requirement of geodesic completeness makes utter sense. Geodesics correspond to the universal free-falling motion of idealized (point-like) time-like observers, and are a natural consequence of the weak equivalence principle imbued in GR and in most extensions of it, which demands the trajectories of all such particles to be independent of their internal composition and structure. This way, the viewpoint above simply hovers the idea that in a physically reasonable space-time all time-like (physical observers) and null (light rays) should not be allowed to cease to exist suddenly or to pop up out of nothingness. Otherwise, the principles of causality, predictability and determinism would be seriously jeopardized. Note, however, that a geodesic completeness check of all trajectories within a given space-time does not guarantee its regularity: it could still harbour other types of pathologies affecting the structure of observers propagating on it. In other words, geodesic completeness may be regarded as a {\it minimum} necessary condition for a (black hole) space-time to have a chance of being regular, working on a separated and an unrelated logic as that of curvature scalars.

Working with geodesic completeness has yet another advantage: the fact that it is the core concept built in the theorems on singularities. These may be formulated under different combinations of hypothesis, but all of them resort to geodesic completeness as the marker of something going ill in the underlying structure of space-time \cite{Curiel,Senovilla:2014gza}.  For the sake of this chapter, we shall adhere to Penrose's formulation of these theorems.

\begin{svgraybox}
{\bf Penrose's 1965 singularity theorem (null version)} \cite{Penrose1965}: If a given maximally extendible space-time $(M,g_{\mu\nu})$ satisfies the following conditions:
\begin{enumerate}
\item $M$ contains a  closed future-trapped null surface, i.e., a two-dimensional compact embedded sub-manifold surface $S$  such that the two families of light rays emerging orthogonally from $S$ to future infinite have a convergent expansion.
\item $M$ admits a non-compact, connected Cauchy hypersurface, i.e., an achronal hypersurface which is met once and only once by all causal geodesics.
\item The null congruence condition, $R_{\mu\nu}n^\mu n^\nu \geq 0$, with $n^\mu n_{\mu}=0$ a null vector,  holds.
\end{enumerate}
then such a space-time contains at least one incomplete null geodesic at a certain value $u_0$ of its affine parameter $u$. The theorem can be trivially upgraded to its time-like version by considering a time-like vector $u^{\mu}u_{\mu}=-1$ in the third condition, and a minor (technical) refinement of its hypothesis.
\end{svgraybox}

(Very) roughly speaking, the first condition of the theorem states that a black hole is present (we shall not deal here with cosmological singularities), the second that we have a well posed initial data problem, while the third is a technical condition that guarantees the focusing of geodesics at certain value of the affine parameter $u=u_0$ and the development of a singularity. The latter condition is equivalent, via the Einstein equations, to the (in the null case) NEC, therefore linking the singular/regular character of the space-time to fundamental lab-verified features of the matter fields threading it. Central to the proof of the theorems is the idea of {\it maximally extensible} space-time, i.e., a space-time that cannot be further extended beyond the focusing point $u_0$, this assumptions designed to avoid cheating the theorems out by considering singular space-times that have made them so by artificially removing a given region of them to provoke the incompleteness of geodesic paths (e.g., Minkowski space with a single point removed). In other words, a geodesic path is non-defined beyond $u_0$ because {\it there is literally no further space it could be occupied by it}.  In this part of the chapter, we shall stick to this idea and check the geodesic (in)completeness of (some of) the black hole solutions introduced in the previous sections.

In facing the issue of building the geodesic equations of our Palatini theories, one could question which one of the two metrics, the auxiliary one, $q_{\mu\nu}$, or the space-time one, $g_{\mu\nu}$, should be considered? We point out that in our construction of RBGs, as given by the action (\ref{eq:RBGaction}), we did not allow the auxiliary metric to couple to the matter sector of the theories. This condition guarantees that the matter fields will follow the laws of motion dictated by the space-time metric, while the auxiliary metric's role is reduced to induce a deformation of the geometrical background in which a given observer carries out its geodesic motion, which is consistent with the fact that it can be integrated out in favour of additional matter fields (seeing the theory as GR with non-linear couplings). Under these conditions, the geodesic equations retains its usual form
\begin{equation} \label{eq:geogen}
\frac{d^2 x^\mu}{du^2}+\Gamma_{\alpha\beta}^{\mu}\frac{dx^\alpha}{du} \frac{dx^\beta}{du}=0
\end{equation}
where $\Gamma_{\alpha\beta}^{\mu}$ are the components of the connection computed with the usual Christoffel symbols of the space-time metric $g_{\mu\nu}$. The Hamiltonian formulation of a point-particle action in a spherically symmetric space-time of the form $ds_g^2=-C(x)dt^2+B(x)dx^2+r^2(x)d\Omega^2$ yields two conserved quantities, namely, the energy per unit mass, $E=C(x)\dot{t}$, and the angular momentum per unit mass, $L=\dot{\varphi}r^2(x)\sin^2 \theta$, where dots denote derivatives with respect to the affine parameter $u$. By the freedom granted by spherical symmetry one can take the motion to be confined to the plane $\theta=\pi/2$ without loss of generality, in such a way that the geodesic equation (\ref{eq:geogen}) can be written in this background as
\begin{equation} \label{eq:geosss}
\frac{C(x)}{B(x)}\left(\frac{dx}{du}\right)^2=E^2-V(x)
\end{equation}
which is somewhat akin to the equation of motion of a single particle moving in a one-dimensional effective potential of the form
\begin{equation}
V(x)=C(x)\left(\frac{L^2}{r^2(x)} -k \right)
\end{equation}
where $k=\{-1,0,+1\}$ corresponds to the sign of the Hamiltonian itself (properly re-scaled) and indicate the nature of the geodesics: time-like $k=-1$ (physical observers), null $k=0$ (photons) and space-like $k=+1$ (tachyons - hypothetical particles moving faster than the speed of light).

Our goal now in verifying the regularity of the space-times considered so far is to inspect every time-like and null geodesic (disregarding space-like trajectories as unphysical) using the equation above, and to see whether each of them can be extended to arbitrarily large values of its affine parameter. Since in our analysis above we found that $f(\mathcal{R})$ theories manifest a qualitatively different behaviour in their metric functions as compared to quadratic/EiBI gravity, we shall consider separately the behaviour of their corresponding geodesics.

For $f(\mathcal{R})$ gravity the line element is given by Eq.(\ref{eq:metricgfR}), so that the geodesic equation (\ref{eq:geosss}) is particularized to this case as
\begin{equation} \label{eq:geogenfR}
\frac{1}{f_{\mathcal{R}}^2}\left(\frac{dx}{du}\right)^2=E^2-V(x) \hspace{0.1cm} ; \hspace{0.1cm} V(x)=\frac{A(x)}{f_{\mathcal{R}}}\left(\frac{L^2}{r^2(x)} -k \right)
\end{equation}
In those cases in which a bounce is present, i.e., when $x^2=z^2/{f_{\mathcal{R}}}$ implements a non-monotonic behaviour in $z(x)$ (recall that this is only possible when $f_{\mathcal{R}}$ has a zero), it is more convenient to rewrite the equation above in terms of $z$ itself as
\begin{equation} \label{eq:uzfr}
\frac{du}{dz}= \pm \frac{\left( 1+\frac{zf_{\mathcal{R},z}}{2f_{\mathcal{R}}} \right)}{f_{\mathcal{R}}^{1/2}\sqrt{E^2+g_{tt}\left(k + \frac{L^2}{r_c^2z^2}\right)}}
\end{equation}
where a factor $r_c$ has also been re-scaled in the affine parameter $u$, while the $\pm$ signs correspond to outgoing/ingoing geodesics. The potentially problematic region in these cases correspond to the location of the bounce itself, $z=z_c$ (as opposed to the RN case, where no bounce is present and one can reach the point $z=0$ instead), so we have to integrate this equation for both time-like and null geodesics, and spanning all possible values of the energy and the angular momentum. Starting with null $k=0$ radial $L=0$ geodesics is our first natural choice, since these are incomplete in the RN geometry of GR. In this case, the above equation reads simply
\begin{equation} \label{eq:geonull}
\pm E \frac{du}{dz}= \frac{\left( 1+\frac{zf_{\mathcal{R},z}}{2f_{\mathcal{R}}} \right)}{f_{\mathcal{R}}}
\end{equation}
so all comes down to the behaviour of the function $f_{\mathcal{R}}$. At large distances, $z \gg 1$, where $f_{\mathcal{R}} \to 1$ (i.e., GR), one finds the expected behaviour there, $Eu \approx \pm z $. On the other hand, at the bounce location $z \to 1$, since $f_{\mathcal{R}} =1-2\sigma \kappa^2 T$, one expects the incompleteness of geodesics as far as this factor goes to a constant by the same reasons as above. This implies that only when the gravity constant $\sigma$ and the trace of the stress-energy tensor have opposite signs will this factor go to zero and one can expect new scenarios of regularity. Indeed, this is exactly the case when the bounce in the radial function is present, since it comes from the relation $x=zf_{\mathcal{R}}^{1/2}$, which requires the presence of a zero in $f_{\mathcal{R}}$.  Therefore, bounce scenarios are precisely those in which one can hope to achieve completeness of this kind of geodesics. Take for instance the quadratic $f(\mathcal{R}) $ case coupled to BI electrodynamics studied in Sec. \ref{sec:1.3.1}. A bounce is present at a minimum value of the radial function given by \cite{Bambi:2015zch}
\begin{equation}
z_c^4(\alpha)=\frac{1+2\alpha-\sqrt{1+4\alpha}}{2\sqrt{1+4\alpha}}
\end{equation}
in such a way that the expansion of the function $f_{\mathcal{R}}$ there becomes of the form $f_{\mathcal{R}} \approx c(\alpha)(z-z_c)+ \mathcal{O}(z-z_c)^2 $ (with $c$ a constant whose shape is irrelevant for our purposes here) and the geodesic equation (\ref{eq:geonull}) thus becomes
\begin{equation}
\pm E (u(z)-u_0) \approx \mp \frac{z_c}{c^{1/2}(z-z_c)^{1/2}}
\end{equation}
where $u_0$ is an integration constant. This expression implies that, as the bounce location $z=z_c$ is approached, the affine parameter diverges to $\pm \infty$. This can be interpreted in terms of ingoing (outgoing) null radial geodesics taking an infinite affine time to get to (depart from) the bounce radius $z_c$; in other words, $z=z_c$ represents the actual boundary of the space-time, which is pushed in these theories to infinite affine distance. Therefore, since the affine parameter can be indefinitely extended in both directions (to asymptotic infinity on one side, and to the infinitely-displaced bounce location on the other), these null radial geodesics are complete in these geometries. This is a strong departure from the GR (RN) result, where the fact that the center of the solution, $r=0$, is reached in finite affine time by these geodesics, without possibility of further extension beyond this point, makes it to be (null) geodesically incomplete, hence singular from this point of view.

What about the remaining geodesics? From (\ref{eq:uzfr}) one can see that either when $k$ or $L^2$ are non-vanishing, the term $g_{tt}$ under the square-root will contribute non-trivially to the geodesic behaviour. Indeed, in the case of BI electrodynamics considered above this term behaves as $g_{tt} \sim -1/(z-z_c)^2$ as $ z \to z_c$. This means that before the surface $z=z_c$ can be reached, the term inside the square root will vanish, and a turning point is reached: in other words, the particle finds an infinite potential barrier preventing any such trajectory from reaching the region $z=z_c$ (similarly as in the usual RN solutions). Therefore, geodesic completeness is achieved for all time-like and null trajectories by the ``trick" of effectively pushing the future (past) boundary of the space-time to unreachable distance to any kind of (causal) observers or information. This same conclusion will be met by any other geometry within the $f(\mathcal{R})$ class coupled to suitable matter fields having qualitative similar behaviours in their metric functions, namely, having a bounce in the radial function and having its metric function $A(z)$ with a suitable divergent behaviour. For instance, both the EH electrodynamics \cite{Guerrero:2020uhn}, the type-I anisotropic fluid (\ref{eq:Gfluid1}) with $\gamma>1$ (in order to support a bounce in the radial function) \cite{Bejarano:2017fgz}, and the type-IV anisotropic fluid (\ref{eq:Gfluid4}) of the same fluid, have a similar functional dependence on the $f_{\mathcal{R}}$ and $g_{tt}$ variables as the BI case, and therefore are also null and time-like geodesic complete.

A different kind of behaviour is met by the type-III anisotropic fluid  (\ref{eq:Gfluid1}), corresponding to the de Sitter cores discussed in the previous sub-section. Indeed, in this case nothing prevents all geodesics to reach the center $r=0$ and, being the curvature perfectly regular, one could tentatively conclude that nothing should prevent these geodesics to continue their path through these cores. However, should we really grant to (regular) curvature the authority to determine when the continuation of geodesic paths is allowed? To analyze this question we turn now to the quadratic/EiBI gravity solutions coupled to Maxwell electrodynamics (in the branch $\epsilon<0$). Particularizing the geodesic equation (\ref{eq:geosss}) to the line element of these theories, Eq.(\ref{eq:quadraticgraveq1}), yields the result
\begin{equation} \label{eq:geogenEiBI}
\frac{1}{\Omega_+^2}\left(\frac{dx}{du}\right)^2=E^2-V(x) \hspace{0.1cm} ; \hspace{0.1cm} V(x)=\frac{A(x)}{\Omega_+} \left(\frac{L^2}{r^2(x)} -k \right)
\end{equation}
Doing the same gymnastics as before, in the null radial case the above equation becomes
\begin{equation}
\frac{1}{\Omega_{+}^2} \left(\frac{dx}{du}\right)^2 =E^2
\end{equation}
For physical insight, it is again more useful to rewrite this equation in terms of the radial function $z(x)$ using the relation of this case $dx/dz=\pm \Omega_{+}/\Omega_{-}^{1/2}$ which yields
\begin{equation}
\pm E du= \frac{dz}{\Omega_{-}^{1/2}}
\end{equation}
and, therefore, in this case all comes down to the behaviour of the function $\Omega_{-}$. Again, difficulties may arise at the bounce location, $z=1$, where the function $\Omega_{-}=1-1/z^4$ has a zero. In any case, the above equation admits an exact integral of the form
\begin{equation}\label{eq:nullradial2}
\pm E \cdot u(x)=\left\{ \begin{tabular}{lr} ${_{2}{F}}_1[-\frac{1}{4},\frac{1}{2},\frac{3}{4};\frac{r_c^4}{r^4}]  z$ & \text{ if } $x\ge 0$ \\
\text{ }\\
$2x_0- {_{2}{F}}_1[-\frac{1}{4},\frac{1}{2},\frac{3}{4};\frac{r_c^4}{r^4}]  z$ & \text{ if } $x\le 0$
\end{tabular} \right. \ ,
\end{equation}
where $\pm$ corresponds to ingoing/outgoing geodesics in the $x>0$ ($x<0$) regions, while ${_{2}{F}}_1[a,b,c;y]$ is a hypergeometric function and the integration constant $x_0={_{2}{F}}_1[-\frac{1}{4},\frac{1}{2},\frac{3}{4};1] =\frac{\sqrt{\pi}\Gamma[3/4]}{\Gamma[1/4]}\approx 0.59907$ comes from matching the expressions of the asymptotic regions $x \gtrless 0$ across the surface $x=0$. This expression clearly shows that the two regions on both sides of the bounce, $x>0$ and $x<0$, can be continuously connected across it, therefore guaranteeing the extensibility of these geodesics to arbitrarily large values of their affine parameter. This behaviour also allows to introduce the canonical interpretation for the nature of the solution, corresponding to the presence of a wormhole structure \cite{VisserBook} consisting of two asymptotically flat space-times $x^+ \in(0,+\infty)$ and $x^- \in(-\infty,0)$, and connected by a throat of non-vanishing areal radius $r^2(x)=r_c^2$ located at $x=0$.

For every other geodesic, the effective potential in (\ref{eq:geogenEiBI}) is non-vanishing. Moreover, given the finiteness of the factor $\Omega_{+}$ everywhere (which takes a value two at the bounce location itself), this problem again corresponds effectively to a classical-mechanical problem of the scattering in a central potential $V(r)$. In turn, this potential is governed by the metric function $g_{tt}$ as given by the expansion (\ref{eq:gtt_series}) near $z=1$, which flips sign depending on whether $\delta_1 \gtrless \delta_c$. This is relevant in order to determine whether a turning point will be present (i.e., an infinity in the potential) or if geodesics will be able to approach the central region of the solutions (at least for some values of their energy). Indeed, for $\delta_1 > \delta_c$ (Reissner-Norstr\"om-like solutions) the potential becomes infinitely repulsive at $x=0$, which makes all these geodesics to be repelled back to asymptotic infinity before being able to reach to $x=0$, while for $\delta_1 <\delta_c$ (Schwarzschild-like solutions) one finds an infinitely attractive potential near $x=0$ instead, making all geodesics with enough energy to overcome the maximum of the potential barrier to unavoidably intersect this point. However, likewise for null radial geodesics above, nothing prevents such geodesics to cross the wormhole throat $x=0$ and propagate through the other asymptotically flat space-time to arbitrarily large values of their affine parameter. And finally, in the case $\delta_1=\delta_c$ (Minkowski-like solutions, since the curvature is everywhere regular in this case), the potential has a certain maximum near $x=0$ and, therefore, depending on its energy every geodesic will be either deflected at a certain distance from $x=0$ or cross it to freely propagate to new regions. Similar features can be found for other matter sources within this theory. For instance, the anisotropic fluid configuration of the EiBI case given by Eq.(\ref{eq:Gz}) has similar solutions as those above for the choices of Type-II $\{s_{\epsilon}=-1,s_{\beta}=+1\}$ and Type-I $\{s_{\epsilon}=-1,s_{\beta}=+1\}$ (with $l_{\epsilon}^2/l_{\beta}^2>1$), with a bounce $z(x)$ and a similar description of the type of horizons as well as the behaviour of geodesics and  of curvature scalars, i.e., finite when $\delta_1=\delta_c^{(\gamma)}$ and divergent otherwise. It is worth pointing out that in such a model there are also de Sitter cores within the Type-II $\{s_{\epsilon}=-1,s_{\beta}=+1\}$ provided that $l_{\epsilon}^2/l_{\beta}^2<1$ and which arise when $\delta_1=\delta_c^{(\gamma)}$, with similar comments as those made in the quadratic $f(\mathcal{R})$ case.

One could naturally wonder whether the presence of a bounce in the radial function is a {\it sufficient} condition for the singularity-removal, given the possibility they allow to extend the geodesics to new regions of the manifold. The answer to this question is negative. Indeed, counter-examples of this are typically given by scalar fields, which are prone to yield geodesically incomplete space-times no matter what their radial functions are doing \cite{Afonso:2019fzv,Guerrero:2021avm}. Moreover, the analysis of the conditions for having geodesically complete solutions are much more obscure than in the cases studied above.

\begin{svgraybox}
From our discussion above, we found that geodesic completeness is achieved via (at least) three different mechanisms (see \cite{Carballo-Rubio:2019nel,Carballo-Rubio:2019fnb} for an extended discussion on this point):

\begin{itemize}

\item The innermost region of the solutions, $x=0$ ($r=r_c$ or $z=1$) is pushed to infinite affine distance, thus being an effective (past/future) boundary of the space-time. Null radial geodesics take an infinite time to get/depart from there, while any other geodesic is repelled back at some distance. Another space-time region may lie beyond $x<0$ ($x>0$) but this is not accessible from the $x>0$ ($x<0$) side. This is the typical scenario in (quadratic)$f(\mathcal{R})$ gravity.

\item The bounce in the radial function allows for the extensibility of (null and time-like) geodesics beyond $x=0$ ($r=r_c$), provided that they are able first to overcome the (maximum of the) potential barrier. This is typically ascribed to quadratic/EiBI gravity.

\item A de Sitter core is formed at $r=0$, where curvature is everywhere regular and geodesics can be extended. They may arise in $f(\mathcal{R})$/quadratic/EiBI gravity.

\end{itemize}
\end{svgraybox}

The bottom line of the discussion above is that the completeness of all null and time-like geodesics can be achieved  in (some of) these space-times, regardless of what curvature is doing. It should be pointed out that the singularity-removal must necessarily come at the expense of the violations of any of the conditions imposed by the singularity theorems. In most approaches to this issue within GR, energy conditions are those paying the prize in order to achieve the de-focusing of geodesics, typically summoning upon quantum-mechanical arguments (either from the matter or the gravitational fields) to save the day. In the Palatini theories of gravity considered in this work, the fact that the field equations can be written in purely Einstein form with a modified stress-energy tensor given by (\ref{eq:Tmunuhat}) implies that one may have scenarios in which the matter field stress-energy tensor derived from the space-time metric satisfies standard conditions while the effective one (sourcing the $q$-formulated Einstein equations in these theories, Eq.(\ref{eq:Tmunuhat})) violates them, so that the null (time-like) congruence condition is bypassed and one can achieve geodesic completeness restoration. This is exactly what happens in most of the geometries considered here, since we chose matter fields satisfying energy conditions from the onset.

\subsection{The non-equivalence between geodesic completeness and curvature divergences}

Our discussion on the extensibility of geodesics (mostly) forsook the issue with curvature divergences. In the $f(\mathcal{R})$ case curvature divergences are typically present, and certainly so in those solutions whose inner region has been pushed to asymptotic infinity. In such a case, the wormhole has only one accessible side, since the other side cannot be reached by observers living in the former at any value of their (endless) affine parameters. Therefore, these curvature divergences should be devoid of any physical meaning since no observer will ever be able to interact with them and experience any type of pathological behaviour. On the contrary, in those models in which a bounce in the radial function is present and accessible to some subsets of geodesics, the presence of curvature divergences at the bounce must be taken seriously. Remarkably, we have seen that there are geodesically complete solutions with divergent curvatures at the bounce (the quadratic/EiBI gravity coupled to some classes of electromagnetic/fluids with $\delta \neq \delta_c$) but also a discrete sub-family of solutions with finite curvature everywhere provided that $\delta_1=\delta_c$. This is quite an unpleasant situation, since any small perturbation to the latter (via e.g. addition of mass or charge) will take the constants of the solution out of this constraint and, therefore, to fall back into the curvature-divergent family. Moreover, since radial null geodesics in those solutions are insensitive to the specific values of the charge and mass, discriminating between them as complete or incomplete based on the behavior of curvature scalars seems rather arbitrary. Should curvature (divergences) have an actual physical meaning {\it in all cases} it would  make little sense that a regular solution could be turned into a singular one by means of any small perturbation. We also point out that in our framework geodesically incomplete solutions with finite curvature (like in the quadratic $f(\mathcal{R})$ case coupled to the Type-II fluid of Eq.(\ref{eq:Gfluid2})) are also possible and, of course, so geodesically incomplete solutions with divergent curvature (e.g. the Type-I fluids of Eq.(\ref{eq:Gfluid1}) with $\gamma<1$) are.

The bottom line of this discussion is the lack of correlation between the (in) completeness of geodesics and the behaviour of curvature scalars. This goes against the initial intuition expressed at the beginning of this section, but agreed with our concern expressed at the conceptual troubles of using curvature scalars as a proxy for supposedly ill-behaviours of the geometry. We next want to take a closer and more physical look at the consequences of having unbound curvature when extended bodies, as opposed to the idealized scenario of geodesic point-like ones, cross regions with such an unbound curvature.

\subsection{Tidal forces and congruences of geodesics}

After all, what is the problem with unbounded curvatures unless they induce any kind of uttermost destructive consequences upon extended bodies?. It turns out that addressing this question can be made using methods developed decades ago, which we now bring forward. We are thus dealing with tidal forces upon extended observers.

\begin{svgraybox}
{\bf Strong vs weak singularities.} A strong (curvature) singularity occurs when all bodies falling into it are crushed to zero volume, no matter what their physical features may be \cite{Ellis:1977pj,Tipler:1977zzb,CK}. This statement, therefore, captures the intuitive idea of space-time singularities as geometric abhorrent phenomena not related to the properties of specific matter configurations. Conversely, a weak singularity occurs when a body has a chance of surviving to an encounter with it. The strong singularity criterion can be enhanced to include some cases where the reference volume remains finite but the body itself undergoes unacceptably large deformations \cite{Nolan:1999tw,Ori:2000fi}.
\end{svgraybox}

Let us start by modelling any such observer by a set of points in geodesic motion but cohesioned by any kind of physical or chemical forces, i.e., a {\it congruence of geodesics}, where local differences in the gravitational field have an impact on the separation between every pair of geodesics as the time goes by. The idea is to define an infinitesimal volume associated to such a congruence and follow its evolution as the potentially problematic region is approached. If such a volume goes to zero, this would entail the crushing of the congruence, i.e., of the observer, hence the presence of a (strong) curvature singularity. Let us label the congruence by a set $x^\mu = x^\mu(u,\xi)$, where $u$ is the affine parameter for every geodesic in the congruence and $\xi$ identifies the specific geodesic under consideration. Thus, the tangent vector for a given geodesic is $u^{\mu}=\partial x^{\mu}/\partial u$ while the Jacobi vector field connecting infinitesimally close geodesics in the congruence is  $Z^\mu = \partial x^{\mu}/\partial \xi$ and satisfies the geodesic deviation equation
\begin{equation}\label{eq:Jacobi}
\frac{D^2 Z^\alpha}{d\lambda^2}+{R^\alpha}_{\beta\mu\nu}u^\beta Z^\mu u^\nu=0 \ ,
\end{equation}
where $D Z^\alpha/d\lambda\equiv u^\kappa \nabla_\kappa Z^\alpha=Z^\beta \nabla_\beta u^\alpha$. The above equation yields six independent Jacobi vectors which are orthogonal to the basis of vectors associated to an adapted orthonormal tetrad parallel-transported along the congruence, $e_a$, as $Z=Z^ae_a$ and $DZ^a/du$. Note that in such a case Eq.(\ref{eq:Jacobi}) allows to find the components of the Jacobi field at any time starting from a reference time $u_i$ as $Z^a(u)={A^a}_b(u) Z^b(u_i)$ (if the three Jacobi fields are zero at $u_i$ then one can define this relation with its derivatives instead). Moreover, every three linearly independent solutions of (\ref{eq:Jacobi}) as $Z_{i}=Z_{i}^a e_a$, $i=1,2,3$ allow  to define the volume invariant as
\begin{equation}
V(u)=\text{det} (Z_{1}^a,Z_{2}^b,Z_{3}^c)
\end{equation}
Therefore one just simply needs to keep track of this volume element after finding a suitable ansatz for the Jacobi fields under the (spherical) symmetry of the system, allowing to solve the equation (\ref{eq:Jacobi}). For instance, in the Schwarzschild case one finds $V(u)\sim u^{1/3}$, which implies its vanishing as $u \to 0$, therefore implying the existence of a crushing-type (strong) singularity alongside the incompleteness of geodesics. On the contrary, in the quadratic/EiBI gravity system coupled to electromagnetic fields (i.e., the geodesically complete solutions with $\delta_1 \leq \delta_c$, since those are the only time-like trajectories able to reach the center of the solutions), as the wormhole throat is approached the infinitesimal volume transported by the congruence goes instead as $V(u) \sim 1/u^{1/3}$ \cite{Olmo:2016fuc}, which diverges as $u \to 0$. Moreover, this happens irrespective of the existence of curvature divergences at the throat, in agreement with our discussion above on the lack of correlation between the behaviour of curvature and the extensibility of geodesics. This result on the volume invariant is not formally included within the standard classification of strong/weak singularities above, so it seems that we have come to a dead end this way.

A more physically-intuitive approach is to consider whether the stretching and subsequent contraction undergone by the reference volume in the congruence as the potentially problematic region is approached and left behind has any physical impact upon every observer crossing it. For instance, in the Schwarzschild solution the radial direction undergoes an infinite stretching while the angular directions undergo an infinite contraction (a process mockingly known as spaghettization). One can capture the heart of this problem by determining whether causal contact among the geodesics in the congruence is lost at any moment. Should the latter happen, then one would conclude that the observer has been destroyed by the presence of a strong curvature singularity. The implementation of this approach in specific scenarios is technically daunting, and so far has only been achieved in the quadratic/EiBI gravity case coupled to a Maxwell field. To make a long story short, one finds that the elements in the congruence never lose causal contact among them despite the formally infinite stretching experienced as the wormhole throat is approached, which is then followed by a rapid contraction as the throat is left behind \cite{Olmo:2016fuc}. In this example, there are no observable destructive effects due to this process, no matter the behaviour (divergent or finite) of the curvature scalars, because the stretching and contraction occurs so fast that it ends up been unobservable.

\subsection{Completeness of accelerated paths}

The principle of general covariance imbued in GR and in most extensions of it dictates the lack of privileges to any set of observers. For our interest here, this means that the completeness of paths of accelerated observers, in addition to the geodesic ones, must be guaranteed for a given space-time to give an accurate description of the physical world. This becomes yet another test for the regularity of a given space-time: to verify that every non-geodesic observer with arbitrary motion and, in particular, with an arbitrary (but bound) acceleration finds infinitely extended paths towards the future and the past. For instance, there are examples (admittedly, toy-like) in which a geodesically complete space-time contains time-like curves with (bound) acceleration but finite total proper length \cite{Geroch:1968ut}, and one would like to make sure that such a thing does not occur in their beloved ``regular" space-times.

Under the presence of external forces, the right-hand side of the geodesic equation (\ref{eq:geogen}) for a massive particle no longer vanishes:
\begin{equation} \label{eq:accmot}
\frac{Du^\mu}{d\tau}=\frac{d^2 x^\mu}{d\tau^2}+\Gamma_{\alpha\beta}^{\mu}\frac{dx^\alpha}{d\tau} \frac{dx^\beta}{d\tau}=a^{\mu}\neq 0
\end{equation}
where $\tau$ is the proper time of a time-like observer. The acceleration undergone by our observer can be computed by considering an orthonormal Frenet-Serret basis $\{u_{(0)},u_{(1)},u_{(2)},u_{(3)}\}$ \cite{Friedman:2015ufz} (parenthesis are introduced here to stress the coordinate-free character of these vectors) to the tangent space satisfying the equation \cite{Olmo:2017fbc}
\begin{equation} \label{eq:FS}
\frac{Du_{(a)}}{ds}=u_{(b)}{A^{(b)}}_{(a)} \hspace{0.2cm};\hspace{0.2cm}
{A^{(b)}}_{(a)}=
\left(
\begin{array}{cccc}
0 &  k(s) & 0 & 0\\
k(s) & 0 & \tau_1(s) & 0  \\
0 & -\tau_1(s) & 0 & \tau_2(s)  \\
0 & 0 & -\tau_2(s) & 0  \\
\end{array}
\right)\ ,
\end{equation}
where the curvature $k(u)$ is interpreted as the linear acceleration experienced by the observer in the direction of $u_{(1)}$, while the first $\tau_1(u)$ and the second $\tau_2(u)$ torsions of the curve correspond to the rotational acceleration $\sqrt{\tau_1^2+\tau_2^2}$ along the axis given by the vector $\omega=\tau_2 \lambda_{(1)}+\tau_1\lambda_{(3)}$. Obviously, when there is neither linear acceleration nor torsion, Eq.(\ref{eq:accmot}) recovers the usual geodesic equation. In a spherically symmetric space-time written in suitable coordinates as $ds^2=-A(y)dt^2+A^{-1}(y)dy^2+r^2(y)d\Omega^2$, making use of the conservation of energy and angular momentum, and restricting ourselves to linear accelerations only for simplicity,  Eq.(\ref{eq:accmot}) can be written, after a straightforward algebraic exercise, as \cite{Olmo:2017fbc}
\begin{equation}\label{eq:acceL}
(u^y)^2+A(y)\left(1+\frac{L^2}{r^2(y)}\right)=\left[E+\int^y_{y_0}\frac{k(s)dy'}{\sqrt{1+\frac{L^2}{r^2(y')}}}\right]^2 \ .
\end{equation}
where $u^y=dy/du$ and $k(y)$ is the total acceleration of the curve in this simplified model. This is the natural extension of the Eq.(\ref{eq:geosss}) to accelerated observers, covering every space-time considered in this chapter. We can thus check the behaviour of any such set of accelerated observers in any pick of geodesically complete sample found above, and for convenience we shall consider (once again) the quadratic/EiBI gravity by its agreeable properties. By considering a test charged (Maxwell) particle which experiences an acceleration (in modulus) of size $k(y)=\frac{Qq}{m r^2(y)}$, and switching by convenience to the variable $x$ defined by the relation $dy=dx/\Omega_+$, then  Eq.(\ref{eq:acceL}) becomes
\begin{equation}\label{eq:beyondGeoBI}
\left(\frac{u^x}{\Omega_+}\right)^2+A(x)\left(1+\frac{L^2}{r^2(x)}\right)=\left(\tilde{E}+I^{BI}_L(r)\right)^2 \ ,
\end{equation}
where $\tilde{E}=E-I^{BI}_L(r_i)$, with $r_i$ the initial location of the particle and the function
\begin{equation}\label{eq:IBIL}
I^{BI}_L(r)=\frac{Qq}{m}\int^r \frac{dr}{r\Omega_-^{\frac{1}{2}}\sqrt{r^2+L^2}}=-\frac{Qq}{m} \frac{\text{EllipticF}\left[\sin ^{-1}\left(\sqrt{\frac{L^2+r_c^2}{L^2+r^2}}\right),\frac{L^2-r_c^2}{L^2+r_c^2}\right]}{\sqrt{r_c^2+L^2}}\ .
\end{equation}
encodes the acceleration term in this scenario. Let us recall the constant character of the radial function at the bounce $x=0$ ($r=r_c$), which in turn makes the function $I^{BI}_l(r_c)$ to become a constant there and, therefore, the single effect induced by the acceleration of our point-like charge is to shift its total energy, therefore having no effect in the extensibility of geodesics. We thus conclude the completeness of all null and time-like paths (in the latter both geodesic and accelerated) in these geometries.

Let us note that in those cases in which the bounce lies at the future (past) boundary of the space-time (quadratic $f(\mathcal{R})$ case coupled to different fluids), one expects that the introduction of bounded accelerations will not alter the conclusion on every time-like geodesic being bounced back by the infinite potential barrier. And what if, by the sake of curiosity, we consider an unbounded acceleration? In such a case, a similar computation as above reveals that such observers would be able to get to the bounce location and gain access to the other side of the wormhole. Such scenarios are, of course, unlikely to be physically achievable.

\subsection{Further tests?}

Up to now we have considered tests with idealized point-like particles (completeness of geodesic and of accelerated paths), and the effects of tidal forces from unbound curvatures upon extended observers. Is there any other further test one could bring to the table?. There is, of course, the important and physically relevant case of the interaction of test waves with potentially pathological space-time regions. In a regular geometry it seems natural to demand that the propagation of any wave through it should be well defined at all times, i.e., that one can establish a well-posed problem of transmission and reflection coefficients. This includes, in particular, the  case of gravitational waves, where we find the notable (and perhaps non-intuitive) fact that in Palatini theories of gravity, these waves propagate on the effective gravitational geometry provided by $q_{\mu\nu}$ rather than the space-time one, see \cite{BeltranJimenez:2017uwv} for details. For the sake of this chapter, however, we shall simplify the problem down to scalar waves propagating upon the latter geometry, a scenario in which relevant insights on the regularity of these space-times can be obtained.

Let us thus consider a massive scalar field with equation of motion $(\Box-m^2)\phi=0$. Using the usual mode decomposition, $\phi_{\omega ml}=e^{-i\omega t}Y_{lm}(\theta)f_{\omega l}(x)/r(x)$, one can find the expression of the equation for its radial component in any space-time of interest. Considering again the quadratic/EiBI gravity case, and formulating the problem in terms of the coordinate $y$ given by $dy/dx=1/(A\Omega_{+})$, one finds a Schr\"odinger-like equation of the form
\cite{Olmo:2015dba}
\begin{equation}
-f_{yy}+V_{eff}(y)f=\omega^2 f \hspace{0.1cm};\hspace{0.1cm} V_{eff}=\frac{r_{yy}}{r} + A(r)\left(m^2+\frac{l(l+1)}{r^2}\right)
\end{equation}
One can thus proceed to define an incoming wave from past null infinity (for horizonless solutions), or from the event/Cauchy horizon (for black holes) and study its behaviour as the bounce location, $r=r_c$, is approached. This problem is better captured by the computation of the transmission and reflection factors, as well as with the cross section of the former. The final result is that no pathologies over these factors, or any other illness of these space-times, are found \cite{Olmo:2015dba}. This implies that, in addition to regular black holes outside their event horizons, which is the relevant case for gravitational wave astronomy, within these space-times one can also consider the propagation of waves over regular naked wormholes, thus naturally connecting the theoretical issue of space-time singularities with the astrophysically-relevant one of gravitational wave imprints from these objects.

\section{Closing thoughts}

Well inside the twentieth-first century, space-time singularities are still a cornerstone upon which many attempts to extend GR are built. The canonical viewpoint within this approach is that space-time singularities are an artifice of the classical GR description, and that an extension of it via additional scalar objects, supposedly emerging as low-energy effective (Planck scale-suppressed) contributions from a quantized version of it, should come to save the day. In this approach one faces the new opportunities offered by a larger flexibility in choosing the main geometrical ingredients imbued in the enlarged theory and, in particular, those associated to the affine connection: curvature, torsion, and non-metricity. In this chapter we have granted the mutually desired divorce to the metric and the affine connection, and considered a theory whose building blocks are made up of these two independent entities. The field equations of the theories constructed this way are thus found by independent variation of the action with respect to each of them, which requires the introduction of workable methods to solve the connection equations, a problem which is missing in the metric formulation. We succeeded at this regard by introducing an auxiliary, connection-compatible metric, in such a way that the field equations can be written as a set of Einstein-like ones for it. This procedure can be systematized for a large class of models built upon scalars out of contractions of the metric with the (symmetric part of the) Ricci tensor, and consequently dubbed them as Ricci-based theories. Revolving around these methods we can solve such field equations either by direct attack, or by using the Einstein frame to construct an equivalent GR action with non-linearly coupled fields, the latter procedure allowing to find new solutions of any RBG theory under consideration using a seed solution from the GR side, via purely algebraic transformations. The latter is dubbed as the mapping method, which should be a basic element in the toolkit of the metric-affine practitioner.

The spectrum of (spherically symmetric) solutions found via any of the two procedures above, is ample both in the gravity and matter sectors. For the former, $f(\mathcal{R})$ gravity, quadratic gravity, and EiBI gravity, are theories extensively studied in the literature (including their higher-dimensional and three-dimensional versions), while in the latter we considered electromagnetic fields (both Maxwell and non-linear), different classes of anisotropic fluids, and scalar fields. We actually found different classes of configurations, from black holes with two and a single (degenerated or not) horizons to naked (horizonless) objects of several types including de Sitter cores. Besides the modified horizon structure, the new gravitational corrections have an impact in the innermost structure, where depending on the gravity/matter combination one may find a bounce in the radial function, which is the most interesting case for the hopes of achieving singularity-removal. Indeed, we identified two main mechanisms for the removal of singularities, namely, either pushing the bouncing region to the future (or past) boundary of the space-time beyond the reach of any geodesics, or by showing that when the bounce is reachable by some sets of geodesics, these can be naturally extended across of it to reach arbitrarily large values of their affine parameter. In the latter case, one can wonder what is then the effect of curvature divergences in those cases where they appear at the bounce, since the extensibility of geodesics is utterly numb to the presence of them. In addition, the de Sitter cores familiar of the GR case, which are free of curvature divergences, come out naturally as a nice surprise within this formalism. Subsequently, we brought to the table four additional criteria as seekers of potential pathologies: the presence of unbound tidal forces upon extended observers, the loss of causal contact upon any two components of a congruence of geodesics, the extension of the geodesic completeness criterion to observers with (bound) acceleration, and tests with (scalar) waves. We extensively discussed these criteria for several of the solutions above, identifying some in which all of them are satisfied, thus representing non-singular geometries from this point of view. Nonetheless, while these are necessary conditions for a space-time to be physically reasonable on its full extent, this does not guarantee that other types of pathological behaviours may arise beyond the limited imagination of the authors. In absence of robust theoretical results telling us when a space-time is {\it regular} rather than when it is {\it singular}, this is how far we have been able to go.

To conclude, different opinions can be found in the community regarding the interpretation and the severity of the difficulties (or lack thereof) with the existence of singularities within GR, but undoubtedly they are a great reservoir of ideas to challenge (even if timorously) Einstein's interpretation of the geometrical ingredients and physical principles underlying the gravitational interaction. In the end, only experiment will tell if regular black holes contain any clue to it.

\begin{acknowledgement}
DRG is funded by the {\it Atracci\'on de Talento Investigador} programme of the Comunidad de Madrid (Spain) No. 2018-T1/TIC-10431. This work is supported by the Spanish Grants  PID2019-108485GB-I00, and PID2020-116567GB-C21, funded by MCIN/AEI/10.13039/ 501100011033 (``ERDF A way of making Europe" and ``PGC Generaci\'on de Conocimiento"), the project PROMETEO/2020/079 (Generalitat Valenciana), the project H2020-MSCA-RISE-2017 Grant FunFiCO- 777740, the FCT projects No. PTDC/FIS-PAR/31938/2017 and PTDC/FIS-OUT/29048/2017, and the Edital 006/2018 PRONEX (FAPESQ-PB/CNPQ, Brazil, Grant 0015/2019). This article is based upon work from COST Action CA18108, supported by COST (European Cooperation in Science and Technology).
\end{acknowledgement}

\bibliographystyle{acm}

\end{document}